\documentclass[twoside]{article}
\usepackage{fancyhdr}
\usepackage{graphicx}
\usepackage{color}
\usepackage{url}
\usepackage{transparent}
\usepackage{amsmath}
\usepackage{hyperref}

\usepackage[italian]{babel}

\definecolor{grey}{rgb}{0.426,0.426,0.426}

\newcommand{\cypher}[1]{\texttt{#1}}
\newcommand{\msg}[1]{\texttt{#1}}
\newcommand{\key}[1]{\texttt{#1}}

\newcommand{\odds}[1]{\textbf{#1}}
\newcommand{\coin}[1]{\textbf{#1}}
\newcommand{\freqlett}[1]{\texttt{#1}}

\newcommand{\n}{\noindent}
\newcommand{\textred}[1]{\textcolor{red}{#1}}

\begin{document}

\vspace*{2.5cm}
\centerline{{\Large Sulla decifratura di Enigma}}
\vspace*{.05cm}
\centerline{{\large Come un reverendo del XVIII secolo contribu\`\i{} }}
\vspace*{.05cm}
\centerline{{\large alla sconfitta degli U-boot tedeschi durante la Seconda Guerra Mondiale}}

\medskip

\centerline{{{\it Fabio S. Priuli \& Claudia Violante}}}

\bigskip

\bigskip

\section{Introduzione}\label{sec:intro}

Con la desecretazione di numerosi documenti risalenti al periodo della Seconda Guerra Mondiale, avvenuta nei primi anni del XXI secolo, si \`e cominciato a comprendere il ruolo determinante giocato da un gruppo di matematici, informatici, enigmisti e scacchisti britannici nella decifratura del ``Codice Enigma'' e quindi nella determinazione dell'esito della Seconda Guerra Mondiale sul fronte occidentale. Nel corso dell'ultimo decennio vi sono stati vari libri e documentari sul tema, e particolare attenzione \`e stata posta sul ruolo decisivo che ebbe il matematico Alan Turing, condannato nel 1952 alla castrazione chimica a causa della sua omosessualit\`a e morto, forse suicida, nel 1954. La riabilitazione di Turing \`e avvenuta solo negli ultimi anni, con le scuse ufficiali da parte del governo britannico nel 2009 e la grazia postuma nel 2013.

Il punto su cui vogliamo fissare la nostra attenzione in questo scritto \`e l'importante ruolo che ricopr\`\i{} nella vicenda un teorema del '700, proposto dal reverendo Bayes per risolvere alcuni problemi sull'equit\`a di lotterie e scommesse. Si tratta della cosiddetta regola di Bayes, che verr\`a presentata e discussa in maggior dettaglio nel paragrafo~\ref{sec:Bayes}: tale regola permette di quantificare correttamente come debba essere aggiornata la valutazione di una probabilit\`a\footnote{Intesa come la misura del grado di fiducia che un individuo coerente attribuisce, secondo le sue informazioni e opinioni, all'avverarsi di un evento (cfr. paragrafo~\ref{sec:Bayes}).} alla luce di nuove informazioni o evidenze.

\n La regola permette quindi di effettuare nel modo pi\`u appropriato il cosiddetto \emph{processo di inferenza}\footnote{Con il termine \emph{inferenza} i filosofi indicano un tipo particolare di pensiero, quello che di solito si chiama {ragionamento} e che si esprime linguisticamente in quella che normalmente si dice una {argomentazione}. Un'inferenza parte da alcune \emph{premesse} e arriva a delle \emph{conclusioni}, ma evidentemente non tutti i modi per compiere tale passaggio sono equivalenti.} per valutare le cause pi\`u probabili che abbiano provocato un fenomeno osservato, pesando opportunamente la probabilit\`a che tali cause avevano prima dell'osservazione (probabilit\`a \emph{a priori}) con la verosimiglianza (o \emph{likelihood}) che il fenomeno osservato sia effettivamente conseguenza della causa considerata. 

\n Proprio in virt\`u di questa specifica capacit\`a di supporto all'inferenza, la regola di Bayes \`e utilizzata ormai da decenni in numerosissime attivit\`a scientifiche ed accademiche (dall'analisi dati nell'ambito dei grandi esperimenti della fisica di frontiera~\cite{LIGO1,LIGO2}, sino alle ricerche in ambito medico~\cite{ClinicBayes} e alle scienze forensi~\cite{Garbolino}), al punto di essere divenuta un requisito indispensabile per la valutazione dei rischi nei progetti NASA~\cite{NASA1, NASA2}. Negli ultimi anni, complice lo sviluppo di algoritmi di previsione ed intelligenza artificiale basati su di essa, le \emph{reti bayesiane}, questa regola ha trovato applicazione con enorme successo anche in progetti di tipo industriale e strategico (dallo sviluppo dei filtri antivirus alla manutenzione predittiva di asset tecnologici, alla previsione dei trend di traffico telefonico).

Riteniamo quindi significativo un approfondimento del ruolo che essa ebbe in un contesto a prima vista inusuale come quello della decifratura dei messaggi tedeschi durante la Seconda Guerra Mondiale. Tuttavia prima di presentare il contributo fornito dal teorema di Bayes dovremo inquadrare il problema della cifratura di messaggi e di come questa si sia sviluppata nel corso dei secoli, per poter cos\`\i{} meglio spiegare la natura delle difficolt\`a incontrate dai crittoanalisti britannici durante il conflitto e per capire l'effettiva utilit\`a che la regola di Bayes ebbe in tale contesto.

\n Nella sezione~\ref{sec:history} presenteremo quindi una serie di metodi di cifratura che sono stati utilizzati sia nell'antichit\`a che pi\`u recentemente, e li sfrutteremo per introdurre alcuni dei concetti chiave della crittografia moderna. Nella sezione~\ref{sec:overviewEnigma} sposteremo la nostra attenzione al contesto storico in cui venne introdotta la macchina Enigma, ossia al periodo tra le due guerre mondiali. Nella sezione~\ref{sec:decifratura} ci concentreremo sulle parti che compongono Enigma, sull'effetto delle varie componenti sulla cifratura dei messaggi, per dedurne alcune delle propriet\`a, e sui metodi di attacco ai messaggi cifrati di Enigma (quello che ai giorni nostri verrebbe chiamato il \emph{reverse engineering} della macchina), con particolare attenzione alle tecniche del \emph{Banburismus} e dello \emph{Scritchmus} ``inventate'' da Turing e dagli altri crittoanalisti di Bletchley Park per decifrare i messaggi di intelligence tedeschi. Infine nella sezione~\ref{sec:FocusBayes} ci concentreremo finalmente sul teorema di Bayes, sul suo utilizzo e sul suo ruolo decisivo nel ridurre il numero di possibili chiavi di cifratura da testare.

\section{Una carrellata storica sui metodi di cifratura}\label{sec:history}

\subsection{Nell'antichit\`a}\label{sec:antique}
La cifratura di messaggi considerati ``strategici'' in modo che solo gli effettivi destinatari possano leggerne il contenuto \`e un problema molto antico che ha visto uomini di grande ingegno proporre soluzioni sempre pi\`u complesse nel corso dei secoli. In questo paragrafo ci limiteremo ad esaminare solo alcune delle soluzioni proposte e storicamente documentate, ovvero quelle che possono essere considerate pi\`u vicine alle moderne macchine cifranti. Ignoreremo quindi le tecniche \emph{steganografiche}, in cui \`e l'intero messaggio a venire celato alla vista per garantirne la riservatezza, come nell'esempio dei messaggi di Demerato di Sparta, narrato da Erodoto nel libro VII delle sue \emph{Storie}~\cite{Erodoto}, in cui le tavolette lignee con inciso il messaggio vennero ricoperte completamente con della cera, per sembrare inutilizzate e non destare il sospetto dei Persiani guidati da Serse. Ci concentreremo, invece, sulle tecniche propriamente \emph{crittografiche}, in cui solo il contenuto viene celato, cos\`\i{} che anche se il messaggio fosse letto da persone non autorizzate, queste non sarebbero in grado di comprenderlo. Presenteremo principalmente esempi in cui la cifratura si basa su \emph{trasformazioni alfabetiche} (in cui cio\`e le singole lettere di un messaggio sono rimpiazzate da altre lettere), tralasciando altre soluzioni quali il \emph{quadrato di Polibio}~\cite{Polibio}, in cui ciascuna lettera del messaggio originale viene sostituita da una coppia di numeri, i quali a loro volta permettono di individuare la posizione della lettera su un'opportuna ``scacchiera'' cifrante.

Possiamo rintracciare le prime documentazioni di tecniche di trasformazione alfabetica nell'antica Grecia, pi\`u precisamente a Sparta durante il IV secolo a.C. Come testimoniato da Plutarco nel suo \emph{Vite Parallele}~\cite{Plutarco}, infatti, gli \emph{efori}, supremi magistrati di Sparta, utilizzavano la cosiddetta \emph{scitala} per inviare messaggi ai propri comandanti militari evitando che altri potessero intenderne il contenuto. Il meccanismo sfruttato era il seguente: una striscia di pergamena veniva avvolta strettamente attorno ad una bacchetta di legno (la \emph{scitala}, appunto) e su di essa veniva scritto il messaggio; dopodich\`e la pergamena veniva srotolata dalla bacchetta ed inviata al destinatario, contenendo all'apparenza soltanto una sequenza di lettere non collagabili tra loro. I destinatari per leggere il messaggio dovevano avvolgere la pergamena su una bacchetta identica, cos\`\i{} che le lettere si allineassero correttamente a formare le parole originarie. In termini moderni si pu\`o rappresentare  il meccanismo della \emph{scitala} come segue:
\begin{itemize}
\item si scrive il messaggio in un rettangolo di lato fissato;
\item si inviano le lettere del messaggio procedendo per colonne, invece che per righe.
\end{itemize}
Ad esempio, per trasmettere il messaggio \msg{DOMANIPARTIREMO}, possiamo inserirlo in un rettangolo di base 5:
$$
\begin{array}{c}
\mbox{\cypher{DOMAN}}\\
\mbox{\cypher{IPART}}\\
\mbox{\cypher{IREMO}}
\end{array}
$$
ed inviare le lettere nel seguente ordine \msg{DIIOPRMAEARMNTO}. Il ricevente, per ricostruire il contenuto del messaggio, dovr\`a scrivere le lettere ottenute dividendole in 5 gruppi:
$$
\begin{array}{c}
\mbox{\cypher{DII}}\\
\mbox{\cypher{OPR}}\\
\mbox{\cypher{MAE}}\\
\mbox{\cypher{ARM}}\\
\mbox{\cypher{NTO}}
\end{array}
$$
e leggere quanto ottenuto per colonne.

Per cominciare a familiarizzare con alcune delle definizioni che verranno utilizzate nel seguito, introduciamo i concetti di \emph{simmetria della cifratura} e quello di \emph{chiave di cifratura} del messaggio. Una procedura di cifratura, sia essa meccanica o algoritmica, si dice \emph{simmetrica} se le operazioni di decifratura non richiedono alcuna informazione aggiuntiva rispetto a quelle per la cifratura. Ad esempio, nel caso appena descritto, una volta noto il numero di lettere alla base del rettangolo usato per cifrare (numero sul quale mittente e destinatario devono accordarsi prima dello scambio di messaggi) il destinatario non ha bisogno di utilizzare informazioni aggiuntive per ricostruire il contenuto della comunicazione. Per quanto riguarda il concetto di \emph{chiave di cifratura}, si utilizza questa espressione per indicare l'insieme di tutti i dettagli, configurazioni e parole che sono necessari al processo di cifratura, in aggiunta alla conoscenza della procedura usata per cifrare/decifrare. Nell'esempio della \emph{scitala} si pu\`o considerare la bacchetta stessa come chiave; nella sua generalizzazione la chiave \`e il numero di lettere che costituisce la base del rettangolo usato per la cifratura (nel nostro esempio, \key{5}).
Per sistemi di cifratura simmetrica, una volta nota la chiave e scelta la procedura, il messaggio pu\`o essere facilmente cifrato cos\`\i{} come decifrato.

Un secondo esempio storico \`e quello dei cosiddetti codici \emph{Atbash} e \emph{Albam}, utilizzati nell'Antico Testamento (libro di Geremia). Il codice \emph{Atbash} prescrive di rimpiazzare la prima lettera dell'alfabeto ebraico (\emph{aleph}) con l'ultima (\emph{taw}), la seconda con la penultima, e cos\`\i{} via. Volendo adattare il processo all'alfabeto latino moderno, esso consiste nell'applicare la seguente \emph{trasformazione monoalfabetica} ossia nel rimpiazzare ciascuna lettera dell'alfabeto con una differente:
$$
\begin{array}{c}
\mbox{\cypher{A} con \cypher{Z}}\\
\mbox{\cypher{B} con \cypher{Y}}\\
\mbox{\cypher{C} con \cypher{X}}\\
\mbox{\cypher{D} con \cypher{W}}\\
\ldots
\end{array}
$$
Per visualizzare pi\`u facilmente la trasformazione risultante ricorriamo alla seguente rappresentazione (che utilizzeremo spesso nel seguito):
$$
\begin{array}{c}
\mbox{\cypher{A B C D E F G H I J K L M N O P Q R S T U V W X Y Z}}\\
\mbox{\cypher{Z Y X W V U T S R Q P O N M L K J I H G F E D C B A}}
\end{array}
$$
in cui abbiniamo a ciascuna lettera dell'alfabeto originario (prima riga) la corrispondente lettera dell'alfabeto criptato (seconda riga). Volendo quindi cifrare con questo metodo la parola \msg{BABILONIA}, otterremmo la parola cifrata \msg{YZYROLMRZ}. Per decifrare un messaggio criptato in questo modo \`e sufficiente riapplicare la cifratura, cosicch\'e ad esempio \msg{YZYROLMRZ} torna ad essere cifrata in \msg{BABILONIA}.
Il codice \emph{Albam}, invece, prescrive di dividere l'alfabeto ebraico in due parti e di rimpiazzare la $n$-esima lettera della prima parte con la $n$-esima lettera della seconda, e viceversa. Riportando all'alfabeto latino moderno tale procedura, possiamo affermare che essa \`e equivalente alla procedura descritta dalla rappresentazione:
$$
\begin{array}{c}
\mbox{\cypher{A B C D E F G H I J K L M N O P Q R S T U V W X Y Z}}\\
\mbox{\cypher{N O P Q R S T U V W X Y Z A B C D E F G H I J K L M}}
\end{array}
$$
Anche questi due esempi ricadono nella famiglia delle cifrature simmetriche e mentre nel primo caso si pu\`o parlare di una procedura senza chiave (in quanto il solo algoritmo di cifratura \`e necessario per la cifratura e la decifratura), nel secondo si pu\`o considerare come chiave della cifratura il numero \key{13}, visto che ciascuna lettera del messaggio originale \`e rimpiazzata dalla lettera che si trova a 13 posizioni di distanza nell'alfabeto.

Un ultimo celebre esempio di crittografia nell'antichit\`a \`e il metodo di cifratura che viene attribuito a Giulio Cesare. Secondo quanto scrive Svetonio~\cite{Svetonio}, infatti, tanto Giulio Cesare quanto suo nipote Augusto, utilizzavano frequentemente, per nascondere il contenuto di alcuni messaggi quello che viene oggi chiamato \emph{Cifrario di Cesare}. Utilizzando la nomenclatura introdotta in precedenza essi applicavano una trasformazione monoalfabetica a tali messaggi, rimpiazzando ciascuna lettera con una differente lettera presa ad una distanza fissata da quella originale. Per esempio, un possibile cifrario di Cesare (che diremo di ``passo 4'') consiste nell'applicare la seguente trasformazione monoalfabetica:
$$
\begin{array}{c}
\mbox{\cypher{A} con \cypher{D}}\\
\mbox{\cypher{B} con \cypher{E}}\\
\mbox{\cypher{C} con \cypher{F}}\\
\mbox{\cypher{D} con \cypher{G}}\\
\ldots
\end{array}
$$
e cos\`\i{} via sino a giungere alla trasformazione di \cypher{Z} in \cypher{C} (in quanto, una volta giunti al termine delle lettere disponibili, la lettera successiva viene cifrata in \cypher{A}). Nella rappresentazione ``compatta'' introdotta per i codici \emph{Atbash} e \emph{Albam}, avremo per tale cifrario:
$$
\begin{array}{c}
\mbox{\cypher{A B C D E F G H I J K L M N O P Q R S T U V W X Y Z}}\\
\mbox{\cypher{D E F G H I J K L M N O P Q R S T U V W X Y Z A B C}}
\end{array}
$$
Da questa rappresentazione si nota anche come la cifratura utilizzata corrisponda ad uno \emph{shift} rigido dell'intero alfabeto (in questo caso di quattro posizioni verso sinistra, visto che la prima lettera A viene trasformata nella quarta lettera D). Per fissare le idee, cifriamo una celebre frase di Cesare stesso con tale trasformazione:
$$
\begin{array}{c}
\mbox{\cypher{VENI VIDI VICI}}\\
\mbox{\cypher{YHQL YLGL YLFL}}
\end{array}
$$
Per quanto ai nostri occhi tale metodo possa sembrare banale e poco sofisticato, esso risult\`o assai efficace all'epoca. Altri autori latini (ad esempio Aulo Gellio~\cite{Gellio}) menzionano il fatto che Cesare fosse solito utilizzare anche metodi pi\`u complessi di cifratura, quando la situazione lo richiedesse, tuttavia non ci sono giunte descrizioni specifiche di come questi funzionassero.
Analizzando il cifrario di Cesare dal punto di vista teorico, osserviamo che \`e anch'esso un esempio di cifratura simmetrica e che la chiave della cifratura si riduce in questo caso al numero \key{4}, ossia alla distanza che intercorre tra la lettera originale e quella cifrata. Poich\'e la distanza \`e fissa in un cifrario di Cesare, la chiave pu\`o essere indicata in modo equivalente tramite la lettera cifrata corrispondente alla lettera \cypher{A}, nel nostro esempio \cypher{D}: infatti una volta che sia nota la cifratura della lettera \cypher{A}, la cifratura delle lettere rimanenti si ricava in maniera immediata. Osserviamo anche che la cifratura di tipo \emph{Atbash} non \`e altro che un cifrario di Cesare con chiave \key{13} o chiave \key{N}.
Dovrebbe ora risultare chiaro come sia possibile produrre continuamente nuove cifrature a partire da una qualunque trasformazione (biunivoca) tra le $26$ lettere dell'alfabeto in se stesse, ossia da un qualunque elemento del gruppo delle permutazioni su $26$ elementi $S_{26}$. Vedremo nel paragrafo~\ref{sec:freq}, per\`o, che tutte queste trasformazioni monoalfabetiche non riescono ad offrire elevati livelli di sicurezza perch\'e non camuffano ``abbastanza'' il contenuto del messaggio, prestando il fianco ad una serie di attacchi statistici che ne minano la robustezza.

Concludiamo la nostra panoramica con una versione pi\`u sofisticata dell'idea utilizzata per il cifrario di Cesare. Questa idea si diffuse nel XVI secolo presso la corte di Enrico III in Francia e divenne nota come il \emph{Cifrario di Vigen\`ere}, bench\'e si faccia oggi risalire la sua invenzione all'italiano Giovan Battista Bellaso, che la descrisse in un suo libro del 1533~\cite{Bellaso}, ben 50 anni prima che ne facesse uso Vigen\`ere~\cite{Vigenere}. L'idea di Blaise de Vigen\`ere fu quella di non ricorrere pi\`u ad una singola trasformazione dell'alfabeto nella cifratura, ma di sfruttare diverse trasformazioni per diverse lettere. In termini moderni si trattava di applicare una \emph{trasformazione polialfabetica} al messaggio da cifrare: si sceglievano un certo numero di sostituzioni monoalfabetiche del tipo cifrario di Cesare e si utilizzava una diversa trasformazione per ciascuna lettera, ripartendo poi dalla prima delle trasformazioni se il numero di lettere del messaggio da cifrare era maggiore del numero di trasformazioni prescelto. In questo caso, la \emph{chiave di cifratura} del metodo era data da una sequenza di chiavi, una per ciascuna delle trasformazioni monoalfabetiche di tipo cifrario di Cesare: ad esempio se alla prima lettera si applicava una cifrature con chiave \key{4} (la medesima dell'esempio precedentemente visto) ed alla successiva una cifratura con chiave \key{7}, la chiave del cifrario di Vigen\`ere ottenuto dalla combinazione di questi due cifrari di Cesare sarebbe \key{47} o, in maniera del tutto equivalente, \key{DG} utilizzando la lettera cifrata corrispondente alla \cypher{4} al posto del numero di lettere di cui si deve spostare ciascuna lettera dell'alfabeto. Questa rappresentazione della chiave fornisce anche un semplice meccanismo per memorizzare la chiave di cifratura scelta: invece di ricordare il numero di lettere per cui si deve far scorrere l'alfabeto in corrispondenza di ciascuna lettera del messaggio ``in chiaro'', si pu\`o memorizzare una parola che rappresenti l'insieme delle trasformazioni da applicare. Riprendiamo l'esempio precedente, la cifratura del messaggio \msg{VENI VIDI VICI}, e applichiamo questa volta un cifrario di Vigen\`ere con chiave \key{LUPO}. In altre parole, applichiamo alla prima lettera del messaggio la trasformazione monoalfabetica che cifra \cypher{A} in \cypher{L}, alla seconda lettera la trasformazione che cifra \cypher{A} in \cypher{U}, alla terza lettera la trasformazione che cifra \cypher{A} in \cypher{P}, alla quarta lettera la trasformazione che cifra \cypher{A} in \cypher{O}, e poi ripetiamo le trasformazioni in ordine per le lettere successive. In altre parole si sfruttano i seguenti alfabeti cifranti:
\begin{itemize}
\item per la prima lettera:
$$
\begin{array}{c}
\mbox{\cypher{A B C D E F G H I J K L M N O P Q R S T U V W X Y Z}}\\
\mbox{\cypher{\textred{L} M N O P Q R S T U V W X Y Z A B C D E F G H I J K}}
\end{array}
$$
\item per la seconda lettera:
$$
\begin{array}{c}
\mbox{\cypher{A B C D E F G H I J K L M N O P Q R S T U V W X Y Z}}\\
\mbox{\cypher{\textred{U} V W X Y Z A B C D E F G H I J K L M N O P Q R S T}}
\end{array}
$$
\item per la terza lettera:
$$
\begin{array}{c}
\mbox{\cypher{A B C D E F G H I J K L M N O P Q R S T U V W X Y Z}}\\
\mbox{\cypher{\textred{P} Q R S T U V W X Y Z A B C D E F G H I J K L M N O}}
\end{array}
$$
\item per la quarta lettera:
$$
\begin{array}{c}
\mbox{\cypher{A B C D E F G H I J K L M N O P Q R S T U V W X Y Z}}\\
\mbox{\cypher{\textred{O} P Q R S T U V W X Y Z A B C D E F G H I J K L M N}}
\end{array}
$$
\end{itemize}
ripartendo dalla prima trasformazione per le lettere successive.
Il messaggio cifrato risultante sar\`a quindi:
$$
\begin{array}{c}
\mbox{\cypher{VENI VIDI VICI}}\\
\mbox{\cypher{GYCW GCSW GCRW}}
\end{array}
$$
Ovviamente la scelta di una chiave breve, e lunga quanto le singole parole del messaggio, se da un lato permette una applicazione mnemonica molto semplice della chiave di cifratura, dall'altro rende molto debole la cifratura stessa offrendo numerose regolarit\`a nel messaggio cifrato (ad esempio, si vede subito che le \cypher{V} iniziali e le \cypher{I} finali di ciascuna parola vengono trasformate sempre tramite la medesima cifratura e, di conseguenza, si traducono nelle stesse lettere \cypher{G} e \cypher{W}) che possono suggerire attacchi efficaci ad un osservatore non autorizzato. La scelta di una chiave pi\`u lunga avrebbe sicuramente rafforzato la sicurezza della cifratura anche se \`e la procedura stessa a presentare debolezze simili a quelle del cifrario di Cesare ogniqualvolta che la chiave scelta non sia di lunghezza comparabile alla lunghezza del messagio da trasmettere.
Concludiamo notando che il cifrario di Vigen\`ere rappresenta un ulteriore esempio di cifratura simmetrica.

\subsection{Attaccare un messaggio cifrato}\label{sec:freq}
Cifrature come quelle presentate nel paragrafo precedente possono apparire sicure perch\'e a prima vista il messaggio risultante dalla trasformazione applicata non presenta, in genere, particolari somiglianze con il messaggio originale. Tuttavia non \`e veramente cos\`\i{}, in quanto anche la procedura utilizzata per modificare i messaggi \`e tale da conservare una delle caratteristiche peculiari di tutti i messaggi di senso compiuto in una lingua: la distribuzione delle lettere nelle parole non \`e uniforme come sarebbe in una stringa di lettere generata in maniera casuale, un fatto gi\`a noto agli Arabi sin dal IX secolo d.C.~\footnote{Tradizionalmente si fa risalire questa scoperta al filosofo e matematico arabo Al-Kandi, 801--873 d.C..}.

Ciascuna lingua ha infatti una propria specifica ``impronta'', data dalla frequenza con cui le diverse lettere compaiono nella parole di senso compiuto. In Italiano, ad esempio, le vocali sono molto pi\`u frequenti delle altre lettere (con l'eccezione della \freqlett{u}) ed alcune consonanti quali la \freqlett{q} e la \freqlett{z} sono molto poco frequenti. Mostriamo in figura e in tabella la distribuzione statistica delle varie lettere dell'alfabeto nella lingua italiana.

\begin{figure}[!htb]
\centering
\includegraphics[width=0.45\textwidth]{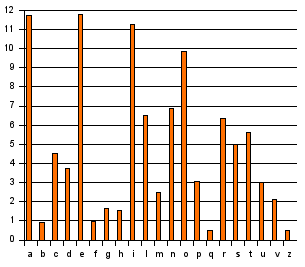}
\caption{Frequenze delle lettere in un testo. (fonte: \emph{Wikipedia}, autore: \emph{Mune$\sim$commonswiki})}
\label{fig:freq}
\end{figure}

\begin{center}
\begin{tabular}{|l|l|l||l|l|l|}
\hline
Lett. & Freq & Lett. & Freq & Lett. & Freq \\
\hline
\hline
\freqlett{a} & $11.74\%$ & \freqlett{h} & $1.54\%$ & \freqlett{q} & $0.51\%$ \\
\freqlett{b} & $0.92\%$ & \freqlett{i} & $11.28\%$ & \freqlett{r} & $6.37\%$ \\
\freqlett{c} & $4.50\%$ & \freqlett{l} & $6.51\%$ & \freqlett{s} & $4.98\%$ \\
\freqlett{d} & $3.73\%$ & \freqlett{m} & $2.51\%$ & \freqlett{t} & $5.62\%$ \\
\freqlett{e} & $11.79\%$ & \freqlett{n} & $6.88\%$ & \freqlett{u} & $3.01\%$ \\
\freqlett{f} & $0.95\%$ & \freqlett{o} & $9.83\%$ & \freqlett{v} & $2.10\%$ \\
\freqlett{g} & $1.64\%$ & \freqlett{p} & $3.05\%$ & \freqlett{z} & $0.49\%$\\
\hline
\end{tabular}
\end{center}
\`E importante osservare che questa distribuzione rimane sostanzialmente invariata se le parole vengono cifrate con il metodo del cifrario di Cesare o con quelli documentati nell'Antico Testamento. 
In questi casi, infatti, a ciascuna lettera \`e associata univocamente la sua ``lettera trasformata'', che rimane sempre la stessa durante tutta la procedura di cirfratura.
Quindi, se si contano le frequenze delle lettere nei messaggi cifrati con una stessa chiave, si potr\`a concludere con una certa sicurezza che tra le lettere pi\`u presenti ci saranno le lettere in cui vengono trasformate la \freqlett{a}, la \freqlett{e} e la \freqlett{i}.
\`E quindi possibile provare a tradurre il messaggio cifrato con queste tre opzioni ed analizzare se sia possibile individuare combinazioni di lettere che suggeriscano parole di senso compiuto oppure provare ulteriori sostituzioni per le consonanti pi\`u frequenti come \freqlett{n}, \freqlett{l} e \freqlett{r}.

Con il cifrario di Vigen\`ere non \`e possibile condurre la medesima analisi delle frequenze, visto che a lettere consecutive corrispondono trasformazioni differenti. Se si fosse tuttavia a conoscenza della lunghezza $L$ della chiave di cifratura e se il messaggio intercettato fosse sufficientemente lungo, sarebbe possibile analizzare separatamente le lettere a cui \`e stata applicata la medesima cifratura, separando il messaggio cifrato in $L$ sequenze separate, ed applicando a ciascun sotto-messaggio l'analisi compiuta in precedenza. In questo modo si potrebbe potenzialmente ricostruire la chiave di cifratura utilizzata e quindi accedere al messaggio originale. 

In generale, ovviamente, la lunghezza della chiave di cifratura non \`e nota ma esistono dei test a cui \`e possibile sottoporre il messaggio intercettato per congetturare quale possa essere la lunghezza della chiave. Il pi\`u celebre \`e quello elaborato da Kasiski~\cite{Kasiski} nel 1863 e consiste nel cercare all'interno del messaggio cifrato delle sequenze di lettere consecutive che si ripetano. A causa della distribuzione non uniforme delle lettere discussa poco sopra, vi \`e una pi\`u alta probabilit\`a che si tratti della medesima sequenza cifrata con la medesima trasformazione che non di due sequenze differenti che casualmente vengono trasformate nella medesima. Quando in effetti questo \`e vero, allora il numero delle lettere che separano le due sequenze deve essere un multiplo della lunghezza della chiave di cifratura, altrimenti non sarebbe la medesima trasformazione ad essere applicata. Se ad esempio la distanza tra le lettere iniziali di tali sequenze fosse pari a $12$, si potrebbe ipotizzare che la lunghezza della chiave di cifratura sia $2$, $3$, $4$, $6$ o $12$. In ciascun caso, scomponendo il messaggio in un numero corrispondente di sotto-messaggi, si pu\`o applicare l'analisi delle frequenze e testare se in effetti vi sia una potenziale corrispondenza tra le lettere nel messaggio cifrato e le lettere di un messaggio in chiaro di senso compiuto.

\subsection{I progressi sino alla fine della Prima Guerra Mondiale}\label{sec:preWWI}
Abbiamo visto sinora che molti dei metodi utilizzati fino al XVII secolo si basano sull'applicazione di una o pi\`u trasformazioni alfabetiche, ossia permutazioni delle lettere, al messaggio da trasmettere e che l'uso di poche trasformazioni (o addirittura di una sola) presta il fianco ad attacchi crittoanalitici basati sull'analisi delle frequenze.
Dovrebbe quindi apparire chiaro che maggiore \`e la lunghezza della chiave di cifratura, ossia il numero delle diverse trasformazioni che vanno a comporre il cifrario completo, maggiore sar\`a la sicurezza garantita. Il numero di lettere che utilizzano la medesima trasformazione, infatti, sar\`a in questo modo minimizzato e render\`a difficile qualsiasi attacco analogo a quelli descritti nel paragrafo~\ref{sec:freq}. 
Portando il ragionamento al limite, si giunge in maniera naturale al cosiddetto \emph{codice Vernam} o \emph{codice One Time Pad} che fornisce una cifratura inattaccabile tramite l'uso di una chiave lunga quanto il messaggio stesso, che non possa essere ripetuta per messaggi successivi. 
\`E interessante notare come il matematico americano Claude E. Shannon, padre della moderna teoria dell'informazione, abbia dimostrato in modo rigoroso l'inviolabilit\`a di questo codice nel 1949~\cite{Shannon}. Eppure, nella pratica, tale codice ``perfetto'' non \`e utilizzabile in situazioni che richiedano scambi frequenti di messaggi tra numerosi utilizzatori. In un simile contesto, infatti, il fulcro del problema si sposterebbe sulla ricerca di una procedura sicura per la trasmissione della chiave stessa, o su come permettere al trasmittente e al ricevente di accordarsi sull'uso di una specifica chiave per la trasmissione del messaggio, e tali problemi sarebbero sostanzialmente equivalenti a quello di partenza.

Senza arrivare a questa soluzione estrema, tuttavia, \`e possibile cercare delle procedure  ``di compromesso'' in cui, pur senza essere garantita una completa inviolabilit\`a, sia fornita una ragionevole sicurezza nelle comunicazioni. \`E sostanzialmente con questo fine che i metodi di cifratura progredirono tra i secoli XVII e XX lungo due diversi binari: da un lato la costruzione di strumenti meccanici che permettessero di applicare con semplicit\`a cifrari di Vigen\`ere con chiavi particolarmente lunghe (citiamo a titolo di esempio il \emph{cilindro cifrante di Jefferson}, proposto dal futuro presidente degli Stati Uniti Thomas Jefferson quando era Segretario di Stato di George Washington, come soluzione per la segretezza delle comunicazioni diplomatiche~\cite{Jefferson}); dall'altro l'utilizzo dei cosiddetti cifrari \emph{poligrafici} o \emph{a repertorio}, soprattutto in ambito diplomatico, in cui si utilizzavano specifici \emph{codebook} che prescrivevano come cifrare specifiche parole chiave ``sensibili'' e poi cifravano il resto del testo suddividendolo in blocchi di 2 o 3 lettere e prescrivendo la regola per cifrare ciascun blocco. 

Questo era lo scenario in cui avvenivano le comunicazioni diplomatiche e militari sino alla fine della Prima Guerra Mondiale ed all'inizio degli anni '20. Allo stesso tempo, per\`o, l'intercettazione dei messaggi cifrati diventava sempre pi\`u facile grazie alla rapida affermazione del telegrafo e della radio come strumenti di comunicazione prediletti: essi da un lato fornivano rapidit\`a e facilit\`a di trasmissione senza pari, ma al contempo erano caratterizzati da una maggiore facilit\`a di intercettazione da parte di ``ascoltatori non autorizzati''. Appare evidente, con il senno di poi, che parallelamente all'ampia diffusione dei nuovi mezzi di comunicazione, non si fosse per nulla affermata una adeguata sensibilit\`a ai rischi legati ad un eventuale furto dei \emph{codebook}. Una simile fuga di informazioni sulle procedure usate da forze armate e corpi diplomatici, congiunta alla semplicit\`a di accesso alle comunicazioni stesse, avrebbe permesso al nemico completo accesso al contenuto delle comunicazioni, concedendogli cos\`\i{} un enorme vantaggio strategico.

Complice di questo scarso interesse verso la sicurezza delle comunicazioni era sicuramente il fatto che alcuni dei grandi successi dell'intelligence inglese durante la Prima Guerra Mondiale erano rimasti secretati anche dopo la fine del conflitto. 
Non era nota, ad esempio, la cattura dell'incrociatore tedesco \emph{Magdeburg} da parte dei russi (nell'agosto 1914) che aveva permesso l'acquisizione dei \emph{codebook} usati dalla Marina tedesca per le loro comunicazioni strategiche. N\`e era noto il recupero del cifrario diplomatico \emph{13040} dai bagagli di un agente tedesco in Persia (nel marzo 1915) da parte di spie britanniche. Tali documenti avevano fornito agli agenti inglesi una finestra spalancata sulle comunicazioni tedesche e ci\`o costitu\`\i{} un chiaro vantaggio per la Gran Bretegna in numerosi episodi. 

\n Un esempio su tutti fu il caso del celebre \emph{telegramma Zimmermann}. Tale messaggio fu inviato nel gennaio 1917 dall'allora Ministro degli Esteri tedesco all'ambasciatore di Germania in Messico, affinch\'e proponesse un'alleanza tra Messico e Germania nel caso gli Stati Uniti fossero entrati in guerra, e venne intercettato dai britannici. Dopo essere stato decifrato, venne passato ai diplomatici USA come leva per convincere il Congresso a dichiarare guerra alla Germania, schierandosi al fianco del Regno Unito nel conflitto.  Furono anche informazioni come questa a giocare un ruolo di primo piano nell'orientare l'opinione pubblica americana verso una posizione interventista, soprattutto dopo che Zimmermann stesso ne conferm\`o la paternit\`a, dissipando i dubbi che si erano diffusi sulla sua autenticit\`a.

Solo dopo il 1923, a seguito della pubblicazione del libro The World Crisis 1911-1918 di Winston Churchill~\cite{Churchill} e di un libro della Marina britannica che raccoglieva i resoconti della Prima Guerra Mondiale dal punto di vista inglese~\cite{NavyWWI}, l'Esercito tedesco venne a conoscenza del danno causato dall'inadeguato livello di sicurezza dei loro messaggi. Fu solo a partire da tale data che si inizi\`o a diffondere la necessaria attenzione verso la sicurezza dei meccanismi di cifratura utilizzati per le comunicazioni pi\`u sensibili.

\n I tedeschi erano comunque in buona compagnia: sino all'inizio della Prima Guerra Mondiale l'Esercito russo neppure cifrava le proprie comunicazioni radio e quello italiano utilizzava una cifratura che era una semplice variante della cifratura di Vigen\`ere (il cosiddetto \emph{cifrario militare tascabile}) di cui erano note le debolezze principali e che infatti poneva una ben misera sfida agli ``uffici delle cifre'' delle altre nazioni europee, i quali potevano facilmente violarne la chiave.

\subsection{Il brevetto di Enigma e le altre macchine cifranti}\label{sec:macchine}
Con l'avvento del XX secolo ed i relativi progressi della tecnica, anche la tecnologia al servizio delle comunicazioni diplomatiche e militari vide un forte sviluppo. Al termine della Prima Guerra Mondiale, un ingegnere tedesco ottenne il brevetto per il progetto di una macchina meccanica capace di cifrare i messaggi tramite un enorme numero di possibili cifrature: Enigma~\cite{Patent}. 
Arthur Scherbius, che aveva compiuto studi di ingegneria presso le universit\`a di Monaco e Hannover, aveva compreso a pieno l'importanza della sicurezza nelle comunicazioni strategiche, sia industriali che militari, e cercava di monetizzare questa sua intuizione proponendo una soluzione tecnicamente molto avanzata e capace di essere adattata sia ad uso civile che militare.

Il \emph{design} era decisamente innovativo, andando a combinare una successione di rotori meccanici, ciascuno capace di applicare una trasformazione monoalfabetica (come descritto nel paragrafo~\ref{sec:antique}), ad alcuni cavi scambiatori, che potevano invertire coppie di lettere del messaggio. Vedremo maggiori dettagli sul funzionamento di Enigma e delle sue componenti meccaniche nel seguito di questo contributo (paragrafo~\ref{sec:EnigmaSet} e soprattutto sezione~\ref{sec:overviewEnigma}), ma per ora basti dire che nessuna macchina per cifrare dell'epoca offriva un livello di complessit\`a della cifratura comparabile. Tuttavia i tempi non erano ancora maturi: quando Scherbius, insieme al suo socio Ernst R. Ritter, propose la sua macchina alla Marina ed al Ministero degli Esteri si sent\`\i{} dire che non c'era interesse ad adottare una simile soluzione. Anche in ambito industriale la risposta fu decisamente fredda e, complice l'alto prezzo a cui era offerta la macchina, gli esemplari prodotti restarono per lo pi\`u invenduti.

\n Solo dopo il 1923, come detto nel paragrafo precedente, si diffuse una maggiore consapevolezza dei danni che comunicazioni insicure possono apportare. Questo cambio di atteggiamento fu particolarmente riscontrabile in ambito militare, laddove i danni erano stati maggiormente evidenti durante il conflitto mondiale, e segn\`o l'inizio del successo di Enigma: nel 1925 ne cominci\`o la produzione su larga scala per l'Esercito e, nei 20 anni successivi, pi\`u di 30000 esemplari furono assemblati e venduti per uso militare e civile in Germania. N\`e si ferm\`o lo sviluppo tecnico: nuovi modelli pi\`u avanzati e sicuri furono adottati durante tutto il periodo di produzione e la richiesta di nuovi modelli, capaci di garantire livello di sicurezza ancora maggiori da parte della Marina, dell'Aviazione e dell'Esercito era pressoch\'e continua.

Per una migliore comprensione del mutato clima nei confronti delle macchine cifranti a partire dalla fine degli anni '20, vale la pena osservare che Enigma non fu l'unica macchina cifrante a venire sviluppata. A seguito del suo successo, infatti, negli anni '30 l'inventore svedese Boris Hagelin svilupp\`o e produsse un gran numero di modelli di machine cifranti,  (tra le tante: la C-36, la C-38 e la M-209), capaci di offrire diversi livelli di sicurezza e diverso ingombro (vi era addirittura un modello portatile, il C-35, con misure pari a circa 13 x 11 x 5 cm). Negli stessi anni i Giapponesi, che durante i negoziati per il Trattato Navale di Washington (1922) erano stati vittime di falle nella sicurezza delle proprie trasmissioni paragonabili a quelle dei tedeschi durante la Prima Guerra Mondiale, cominciarono a sviluppare i primi prototipi di macchine cifranti a rotori (i progetti \emph{Red} e \emph{Purple}, nomi in codice per le \emph{Typewriter Type 91} e \emph{Type 97}, rispettivamente). Anche negli Stati Uniti vennero sviluppate alcune nuove macchine cifranti, quali la macchina a rotori di Hebern o la SIGABA sviluppata durante la Seconda Guerra Mondiale da William Friedman, all'epoca direttore del \emph{Signals Intelligence Service} dell'Esercito degli Stati Uniti. Esse non ebbero un analogo successo a causa della fragilit\`a e dell'ingombro che le caratterizzavano, nonostante fornissero una sicurezza comparabile (addirittura superiore nel caso della SIGABA) a quella di Enigma.

\n Menzioniamo infine il fatto che anche in Germania, accanto alle macchine Enigma, vennero sviluppare durante la Seconda Guerra Mondiale ulteriori macchine cifranti, dotate di un maggior numero di rotori, che venivano collegate direttamente alle telescriventi fornendo un alto livello di sicurezza per le comunicazioni tra l'Alto Comando a Berlino ed i centri di comando dell'Esercito nei vari paesi occupati. Ricordiamo tra queste le macchine cifranti Lorenz SZ40 e SZ42 e le Siemens \& Halske T52, in produzione ed uso tra il 1942 ed il 1945. In particolare le macchine Lorenz furono storicamente rilevanti in quanto portarono allo sviluppo di \emph{Colossus}, il primo computer programmabile basato sull'elettronica digitale, che venne costruito alla GC\&CS di Bletchley Park (il principale centro di ricerca Alleato sulle cifrature, di cui parleremo pi\`u approfonditamente nel paragrafo~\ref{sec:Bletchley}) da un gruppo di ingegneri capitanato da Tommy Flowers~\footnote{Va detto che, diversamente da quanto accade nei computer attuali, la programmazione in \emph{Colossus} consisteva in una serie di connessioni e interruttori da posizionare correttamente sulla macchina stessa e non nella ``compilazione binaria'' di un eseguibile residente nella memoria del calcolatore.}. La decifrazione di queste macchine telescriventi, chiamate in codice \emph{Tunny} dai crittoanalisti, viene anche ricordata in quanto portata a termine sulla base dei soli messaggi intercettati e di ragionamenti probabilistici e crittoanalitici simili a quelli che presenteremo nel seguito, ma senza avere a disposizione alcuna macchina fisica che potesse aiutare a ricostruirne il funzionamento. Rimandiamo il lettore interessato a~\cite{SZ42,Tunny}.

\subsection{Breve descrizione dei settaggi configurabili in una macchina Enigma}\label{sec:EnigmaSet}
Mentre una pi\`u approfondita descrizione della cifratura verr\`a riportata nella sezione~\ref{sec:overviewEnigma}, \`e opportuno fornire a questo punto una prima descrizione del funzionamento della macchina Enigma e, in particolare, dei settaggi quotidianamente a disposizione degli operatori.

A prima vista una macchina Enigma pu\`o sembrare una specie di macchina da scrivere o telescrivente, essendo dotata di una tastiera che permetteva la digitazione delle $26$ lettere dell'alfabeto (i numeri venivano scritti per esteso nelle comunicazioni, ossia \msg{EINS}, \msg{ZWEI}, \msg{DREI}, ecc.) ed un pannello luminoso su cui l'operatore poteva leggere la lettera cifrata corrispondente a ciascuna lettera in chiaro del messaggio. Non vi era modo di stampare il messaggio cifrato e quindi si doveva prendere nota delle lettere del messaggio una per volta, mano a mano che le lettere originali venivano digitate.

La cifratura che una macchina Enigma applicava al messaggio si compone sostanzialmente di due parti: una \emph{trasformazione polialfabetica}, applicata dai rotori, ed uno scambio tra un certo numero di coppie di lettere, applicato tramite i cavi della \emph{plug board}. 

\n La seconda componente \`e la pi\`u semplice da descrivere: ogni giorno l'operatore veniva istruito su quali coppie di lettere dovessero essere scambiate dopo l'applicazione della prima cifratura da parte dei rotori, ed egli inseriva i corrispondenti cavi scambiatori nella \emph{plug board}  frontale della macchina, ottenendo gli scambi desiderati.

\n Il cuore della cifratura per\`o risiedeva nei suoi rotori: tali rotori non erano altro che dei dischetti rotanti con una serie di contatti elettrici su entrambi i lati; i cavi che connettevano i contatti all'interno dei dischi implementavano una permutazione ``fissa'' delle $26$ lettere dell'alfabeto, ossia una trasformazione monoalfabetica di poco pi\`u complessa rispetto a quelle descritte nel paragrafo~\ref{sec:antique}. Ci\`o che rendeva la cifratura ottenuta tramite le macchine Enigma cos\`\i{} sicura era il fatto di utilizzare una trasformazione diversa per ciascuna lettera del messaggio da inviare, senza bisogno di organizzare scambi di chiavi di cifratura troppo complesse. Meccanicamente si otteneva tale effetto montando tre rotori in serie, interconnessi come pezzi adiacenti di un meccanismo, con il terzo rotore che avanzava di una posizione ogniqualvolta una lettera veniva cifrata. Dopo un certo numero di scatti, il terzo rotore provocava un avanzamento nel secondo rotore, e questi a sua volta poteva fare avanzare il primo rotore quando raggiungeva un certa posizione. In questo modo l'effetto complessivo di una macchina Enigma \`e quello di una trasformazione polialfabetica applicata all'intero messaggio, con una chiave lunga $(26)^3= 17576$ caratteri, ossia una cifratura di grande sicurezza: la medesima trasformazione non potrebbe mai venire applicata due volte ad uno stesso messaggio, se non dopo una rotazione completa di tutti e tre i rotori, e nessun operatore si sognerebbe di inviare messaggi lunghi pi\`u di $17000$ caratteri in un'unica trasmissione.

Ad aggiungere ulteriore complessit\`a vi erano il fatto che la serie di trasformazioni applicate al messaggio dipendeva dalla posizione iniziale dei tre rotori, che usualmente mutava per ciascun messaggio, ed il fatto che le permutazioni ``fisse'' applicate da ciascun rotore potevano essere modificate tramite un ``anello'' che correva lungo il dischetto rotante. Con questo ulteriore accorgimento era possibile ``traslare'' di un certo numero di posizioni i contatti elettrici, e quindi di ``traslare'' parimenti la cifratura applicata dello stesso numero di lettere.

\n Riassumendo, quindi, ogni giorno i settaggi della macchina Enigma riguardavano: 
\begin{itemize}
\item la scelta dei rotori da utilizzare e del loro ordine (\emph{Walzenlage}), 
\item la scelta della posizione dell'anello di ciascun rotore (\emph{Ringstellung}), 
\item la scelta delle posizioni iniziali in cui porre i rotori (\emph{Grundstellung}),
\item infine la scelta delle coppie di lettere da scambiare tramite i cavi della \emph{plug board}  (\emph{Steckerverbindungen}). 
\end{itemize}

\n Le versioni della macchina utilizzate negli anni '20 usavano solo tre rotori, il cui ordine veniva scelto dagli operatori ogni giorno, e sei cavi scambiatori. Le versioni evolute utilizzate a partire dai tardi anni '30 invece, permettevano di scegliere i tre rotori tra cinque a disposizione e fornivano dieci cavi scambiatori. La Marina tedesca, ossia la \emph{Kriegsmarine}, era fornita di un numero maggiore di rotori, ben 8, tra cui venivano scelti i tre da utilizzare in una certa giornata. A partire dal 1942, la \emph{Kriegsmarine} ottenne anche la costruzione di una versione ulteriormente evoluta in cui i rotori utilizzati in serie erano quattro invece che tre, con un quarto rotore che poteva essere scelto solo tra due a disposizione e che non ruotava, ma poteva essere configurato come gli altri in una qualunque delle $26$ posizioni iniziali possibili.

Nel seguito del nostro scritto, concentreremo la nostra attenzione per lo pi\`u sulla versione di Enigma a tre rotori e 10 cavi scambiatori, con i rotori scelti tra cinque disponibili, come avveniva per la maggior parte delle comunicazioni tra il 1939 ed il 1942. Osserviamo anche che in questo paragrafo, pur analizzando una larga parte delle componenti di Enigma, abbiamo trascurato una serie di dettagli tecnici, come la presenza del riflettore accanto ai rotori o la frequenza con cui ciascun rotore provoca l'avanzamento del rotore a lui adiacente. Torneremo ad analizzarli nel seguito (pi\`u precisamente nella sezione~\ref{sec:overviewEnigma}) quando descriveremo pi\`u specificamente le propriet\`a della cifratura che potevano essere ricavate a partire dalla conoscenza del funzionamento meccanico di Enigma.

\subsection{Il periodo tra le due guerre}\label{sec:betweenWars}

Con l'adozione della macchina Enigma da parte dei diplomatici e militari tedeschi per cifrare le proprie comunicazioni, i servizi di intelligence britannici e francesi si trovarono alle prese con un problema ben pi\`u complesso di quelli che avevano dovuto affrontare sino alla Prima Guerra Mondiale. Nonostante la possibilit\`a di avere accesso ad alcune macchine Enigma di tipo civile e commerciale, infatti, divenne presto ben chiaro che l'alto numero di possibili configurazioni della macchina rendeva proibitivo il lavoro di decifrazione dei messaggi intercettati.

Sino alla fine degli anni '20, quindi, i servizi di intelligence della maggior parte delle nazioni europee ritennero semplicemente inviolabile la cifratura offerta da Enigma, complice anche il minor senso di urgenza percepito nel periodo di relativa pace e stabilit\`a. Un diverso sentimento era per\`o provato dai polacchi che, schiacciati tra le due grandi potenze tedesca e russa a seguito degli accordi stipulati al termine della Prima Guerra Mondiale, nutrivano un grande interesse strategico verso una effettiva decifrazione dei dispacci tedeschi. Tale interesse si tradusse, nel periodo a cavallo tra la fine degli anni '20 e i primi anni '30, in ingenti investimenti sulla ricerca di nuovi attacchi crittoanalitici che potessero incrinare la sicurezza delle comunicazioni tedesche, accompagnati dalla creazione di un intero dipartimento governativo dedicato allo scopo: il \emph{Biuro Szyfr\'ow}. 

\n Si pu\`o dire che si tratt\`o di denaro ben speso visto che, a fine 1932, i tecnici polacchi del gruppo guidato da Marian Rejewski, Jerzy R\'o$\dot{\rm z}$ycki e Henryk Zygalski riuscirono effettivamente a violare la cifratura delle macchine Enigma a tre rotori usate all'epoca, tramite una combinazione di ragionamento matematico e attacchi \emph{brute force} condotti da macchine meccaniche dette ``\emph{bomba kryptologiczna}''. Si trattava sostanzialmente di un approccio in due fasi: dapprima si sfruttavano le informazioni disponibili sulla struttura dei messaggi trasmessi, cos\`\i{} da ridurre il numero di possibili settaggi che fossero compatibili con i messaggi intercettati; al termine di questo primo lavoro di riduzione, tutte le combinazioni rimaste venivano testate in maniera sistematica dalle macchine \emph{bomba}, fino a giungere all'effettiva individuazione delle configurazioni giornaliere della macchina e quindi alla decifrazione dei messaggi.

Questo successo polacco, presto condiviso con gli omologhi servizi francesi loro alleati, permise di avere nuovamente accesso alle comunicazioni tedesche per gran parte degli anni '30. A partire dal 1938, per\`o, i tedeschi passarono ad una versione pi\`u evoluta di Enigma che, pur utilizzando ancora tre rotori per la cifratura del messaggio, ne metteva a disposizione cinque in totale permettendo cos\`\i{} ai tedeschi di scegliere giornalmente tra ben 60 combinazioni dei rotori (che rappresentano i possibili modi di ordinare tre oggetti distinti a partire da cinque possibili) invece dei 6 originali, rendendo la complessit\`a della decifrazione superiore a quella affrontabile dalle macchine \emph{bomba} costruite dagli analisti polacchi.

\subsection{Bletchley Park e lo scoppio della Seconda Guerra Mondiale}\label{sec:Bletchley}
Allo scoppio della Seconda Guerra Mondiale nel 1939, i servizi di intelligence di tutto il mondo si trovarono sostanzialmente impreparati di fronte alla sfida posta dai messaggi cifrati con i modelli pi\`u avanzati di Enigma. Pur essendo in contatto con l'intelligence polacca, e quindi potendo fare tesoro dei successi ottenuti dai decifratori polacchi, l'elevato numero di settaggi con cui potevano essere configurate quotidianamente le macchine Enigma tedesche era semplicemente non affrontabile con i mezzi a disposizione. Senza contare il fatto che si trattava di una continua gara contro il tempo: al termine di ciascuna giornata le configurazioni della macchina venivano modificate, riportando il lavoro di decifrazione al punto di partenza.

La Gran Bretagna, molto sensibile al tema della decifratura dei messaggi, aveva istituito a fine 1938 la \emph{Government Code and Cypher School} (GC\&CS) a Bletchley Park, nel Buckinghamshire, e dall'autunno 1939 aveva cominciato a reclutare matematici e scienziati per attaccare i messaggi in codice di Esercito, Aviazione e Marina tedeschi.

\n Si trattava di una scelta non ovvia e \emph{all'avanguardia} perch\'e sino ad allora la crittografia era stata vista come un'arte pi\`u consona a letterati ed enigmisti che a scienziati. Tuttavia, i progressi della meccanica e l'avvento delle macchine per la cifratura rendevano necessario un nuovo approccio, pi\`u sistematico e scientifico, da affiancare all'usuale expertise sfruttata sino ad allora.

\n Tra le personalit\`a, reclutate dal governo britannico per lavorare a Bletchley Park e provenienti dai dipartimenti di matematica del Regno Unito, menzioniamo Peter Twinn (dall'Universit\`a di Oxford), John Jeffreys, Alan Turing e Gordon Welchman (dall'Universit\`a di Cambridge) e, negli anni successivi allo scoppio della guerra, Derek Taunt, Irving J. Good, Bill Tutte e Max Newman, reclutati da vari atenei britannici. Accanto a costoro, furono chiamati anche storici come Harry Hinsley, campioni di scacchi come Hugh Alexander e Stuart Milner-Barry, e altri accademici come Dilwyn Knox, esperto di papirologia.

Le attivit\`a frenetiche delle centinaia (e poi migliaia) di persone impiegate nelle \emph{Hut} e nei \emph{Block} di Bletchley Park, ossia i due tipi di strutture a cui i crittoanalisti erano assegnati con compiti diversi, coprivano tutte le fasi dell'intelligence, dall'intercettazione alla decifrazione, alla traduzione dei messaggi decifrati, sino al loro invio all'MI6, ossia alla divisione dell'intelligence militare che si sarebbe occupata dell'utilizzo delle informazioni ottenute e che avrebbe preparato le giustificazioni per le informazioni ottenute. Era infatti fondamentale che i nemici non sospettassero mai che si fosse giunti alla completa decifrazione dei messaggi ed era necessario trovare spiegazioni sufficientemente credibili per le ``fughe di notizie'', quali agenti tedeschi doppiogiochisti che passassero informazioni ad inglesi o francesi, o falle dell'intelligence degli alleati della Germania. L'insieme delle informazioni cos\`\i{} ottenute era definita \emph{Ultra intelligence} perch\'e coperta da un livello di segretezza superiore a quelli usati sino ad allora (che arrivavano sino a \emph{Most Secret}). 

Il maggior risultato degli sforzi dei gruppi di crittoanalisti di Bletchley Park fu lo sviluppo di una versione evoluta delle macchine \emph{bomba} polacche, capace di attaccare anche i messaggi criptati con le macchine Enigma rinnovate. Vi lavorarono per lungo tempo sia Turing che Welchman e altri membri assegnati al team della \emph{Hut 8}. Infatti se da un punto di vista concettuale le macchine polacche erano potenzialmente in grado di trovare le chiavi di cifratura per i messaggi delle nuove macchine Enigma, nella pratica tali macchine \emph{bomba} avrebbero impiegato anni per scoprire la chiave usata dai tedeschi per un singolo giorno, rendendo pressoch\'e inefficace il lavoro di intelligence svolto quotidianamente. Allo stesso tempo mancavano i fondi per costruirne in numero maggiore, per tentare una ``parallelizzazione'' della ricerca, visto che ciascuna macchina costava circa 100.000\textsterling{} e la Gran Bretagna in quel momento poteva difficilmente permettersi grossi investimenti, visti i continui costi da sostenere per contrastare gli attacchi delle forze dell'Asse.

L'unica speranza per decifrare i messaggi tedeschi in tempo utile era quella trovare un modo per ridurre sensibilmente il numero di possibili chiavi da testare, ed \`e in questo campo che il \emph{teorema di Bayes} si rivel\`o infine un ingrediente di enorme importanza. Come vedremo nella sezione~\ref{sec:FocusBayes}, infatti, tale teorema permetteva di scartare come incompatibili con i messaggi intercettati un gran numero di possibili configurazioni giornaliere e questo, congiuntamente con l'utilizzo dei cosiddetti \emph{crib} (su cui torneremo nel paragrafo~\ref{sec:crib}) quando questi erano disponibili, riduceva la mole di lavoro a carico delle macchine \emph{bomba} quanto bastava per far breccia nel muro della sicurezza di Enigma.

\subsection{Mosse e contromosse durante il conflitto}\label{sec:duringWWII}
Durante la guerra ci furono varie evoluzioni dei modelli di macchine Enigma utilizzate, soprattutto per quanto riguarda la versione usata dalla Marina tedesca. Infatti la \emph{Kriegsmarine}, ben conscia del ruolo cruciale giocato dagli U-boot nella guerra di mare, continuava ad affinare le strategie di invio dei propri messaggi e a richiedere versioni ancora pi\`u sicure della macchina.
Per esempio i messaggi non potevano mai essere pi\`u lunghi di un certo numero di caratteri, alcune parole venivano scambiate prima della cifrature con altre, utilizzando specifici \emph{codebook}, e per la cifratura venivano usate macchine Enigma dotate di un numero maggiore di rotori disponibili (i tre rotori erano scelti tra 8 possibili invece che 5). 

L'esistenza di un livello di complessit\`a addizionale nei messaggi della \emph{Kriegsmarine} era noto agli Alleati, che tentarono a pi\`u riprese di recuperare i manuali di utilizzo dagli U-boot attaccati, seppure senza successo. Nel 1940 Ian Fleming, destinato a creare il personaggio di James Bond dopo il conflitto e all'epoca consigliere della \emph{Naval Intelligence Division}, propose di inscenare un incidente aereo con soldati britannici camuffati nei pressi di una nave tedesca per cercare di cogliere di sorpresa l'equipaggio e recuperare i manuali utilizzati dai marconisti tedeschi. L'operazione non venne messa in atto, ma testimonia l'elevato interesse verso tali documenti a tutti i livelli della catena di comando. Gli Alleati riuscirono a mettere le mani su una di queste macchine evolute e sui relativi rotori solo nel maggio 1941, quando l'equipaggio del cacciatorpediniere \emph{HMS Bulldog} riusc\`\i{} a salire a bordo dell'\emph{U-110} e il sottotenente di vascello David Balme riusc\`\i{} a salvare dalla distruzione sia la macchina Enigma che i relativi manuali di configurazione. 

In ogni caso, fu una vittoria di portata limitata: la Marina tedesca aveva gi\`a avviato un piano per rendere ancora pi\`u sicura la propria cifratura, utilizzando una diversa procedura per scegliere le chiavi di cifratura, che sarebbe divenuta effettiva nel novembre dello stesso anno. Tale nuova procedura consisteva nell'applicazione di multiple trasformazioni della chiave giornaliera, spiegate in ``manuali di istruzioni'' che i comandanti di vascelli e sottomarini avevano l'ordine di distruggere in caso di cattura. Per ulteriore precauzione i manuali erano stampati con un inchiostro che scoloriva velocemente a contatto con l'acqua cos\`\i{} da accelerarne il deterioramento in caso di affondamento.
Oltre a questo cambio di procedura, nel febbraio del 1942 la \emph{Kriegsmarine} adott\`o anche una nuova versione di Enigma (a 4 rotori invece che 3). Con queste modifiche i messaggi tornarono ad essere del tutto inattaccabili per pi\`u di dieci mesi.

Allo stesso tempo, per\`o, gli Alleati continuarono ad affondare e catturare U-boot, fino al recupero di una macchina Enigma del nuovo modello nell'ottobre 1942. Tale recupero, reso possibile dalla cattura dell'\emph{U-559} da parte dell'\emph{HMS Petard}, forn\`\i{} gli strumenti necessari ai gruppi di analisti all'opera a Bletchley Park per decifrare le nuove procedure e porre fine al \emph{black out} nell'intelligence navale.

Si trattava sostanzialmente di un inseguimento continuo, in cui ad ogni passo in avanti dei tedeschi gli Alleati dovevano trovare nuove vie per colmare la distanza. Tuttavia, non si trattava di un'impresa disperata: le coperture fornite per giustificare le informazioni fornite dall'intelligence \emph{Ultra} reggevano e la \emph{Kriegsmarine} era sostanzialmente all'oscuro dei progressi fatti dagli Alleati, sentendosi assolutamente sicura dell'inviolabilit\`a delle proprie trasmissioni. Anche al termine della guerra molti ufficiali tedeschi si rifiutarono di accettare la rivelazione che Enigma fosse stata violata: costoro erano assolutamente convinti che i vari casi in cui gli Alleati avevano previsto manovre tedesche e anticipato le contromisure fossero sempre dovuti a casi fortuiti o al tradimento di proprie spie, ma mai si sognarono che a Bletchley Park fossero davvero riusciti a decifrare i messaggi criptati con Enigma. Questo forniva un importante vantaggio agli Alleati, visto che il senso di sicurezza evitava che le macchine Enigma venissero rimpiazzate da versioni pi\`u complesse o addirittura da macchine pi\`u moderne, la cui procedura di decifrazione sarebbe dovuta ripartire da zero.

\subsection{Impatto sulla guerra}\label{sec:conseguenze}
L'impatto delle informazioni intercettate e decifrate tramite l'\emph{intelligence Ultra} sul conflitto fu enorme. Innanzi tutto le informazioni sulla posizione degli U-boot nel Mare del Nord, nella Manica e nell'Atlantico furono cruciali per ridurre le perdite di vite umane tra gli equipaggi degli incrociatori e dei mercantili delle potenze Alleate e per permettere l'arrivo di beni di prima necessit\`a in Gran Bretagna. 

\n Tuttavia vi furono anche numerosi episodi specifici in cui le informazioni ottenute grazie al lavoro degli analisti di Bletchley Park si rivelarono essenziali per l'esito favorevole agli Alleati~\cite{UltraRef1,UltraRef2,UltraRef3}. Ne menzioniamo qui tre tra i tanti: la battaglia di Alam el Halfa, la battaglia di Capo Nord e lo sbarco in Normandia.

Partiamo dal fronte nord-africano dove, nel settembre del 1942, Rommel era da poco stato sconfitto nella prima battaglia di El Alamein e stava pianificando di portare un nuovo attacco alle forze alleate che impedivano l'avanzata verso Il Cairo. In particolare, egli intendeva attaccare alle spalle le truppe in movimento del comandante Montgomery, presso le alture Alam el Halfa. Montgomery, informato dall'intelligence dell'intenzione tedesca di attaccare da sud, decise di lasciare un varco nello schieramento in quella direzione e di attendere le truppe tedesche schierando molti dei suoi carri armati sulle alture. Caduti nell'inganno ed esposti al fuoco dei carri dalla posizione sopraelevata, i tedeschi non poterono far altro che ritirarsi e tornare a difendere le posizioni raggiunte ad El Alamein. Tra l'ottobre ed il novembre successivi El Alamein sarebbe stata teatro della seconda omonima battaglia, in cui le truppe dell'Asse avrebbero riportato una netta sconfitta che avrebbe rappresentato un punto di svolta per lo svolgimento della campagna del Nord-Africa.

Per quanto concerne la battaglia di Capo Nord, la Marina tedesca durante il 1942 e 1943 aveva affondato al largo della Norvegia numerosissimi convogli che cercavano di portare rifornimenti in Russia, sfruttando tra gli altri l'incrociatore \emph{Scharnhorst}. Gli inglesi per cercare di neutralizzare la minaccia rappresentata da tale incrociatore decisero di tendergli una trappola. Un convoglio inglese partito il 20 dicembre e diretto verso la Russia avrebbe cercato di attirare l'attenzione dei tedeschi e la Marina britannica avrebbe cercato di isolare l'incrociatore e neutralizzarlo al largo della costa norvegese, con lo sforzo congiunto delle navi \emph{Destroyer} che accompagnavano il convoglio e di quelle che scortavano un precedente convoglio di ritorno dalla Russia. Nei giorni intorno al 23 dicembre i messaggi intercettati tra la \emph{Kriegsmarine} e la \emph{Luftwaffe} mostravano chiaramente che il convoglio diretto verso la Russia, con i suoi 19 vascelli pi\`u la scorta, aveva attirato l'attenzione dell'Aviazione tedesca e che quindi il piano Alleato poteva procedere come deciso. La notte del 25 dicembre l'incrociatore \emph{Scharnhorst} lasci\`o il porto di Altenfjord, che utilizzava come base operativa, per intercettare i convogli al largo di Capo Nord. Tuttavia, sfruttando le avverse condizioni meteorologiche, che rendevano impossibile all'Aviazione tedesca di proseguire una sorveglianza pressante, gli Alleati deviarono la rotta del convoglio verso nord-est e lasciarono in zona solo le navi della scorta. La mattina del 26 dicembre l'incrociatore \emph{Scharnhorst} non fu in grado di individuare la posizione del convoglio e diede ordine ai cacciatorpedinieri al suo seguito di allargare l'area di ricerca, separandosi dallo \emph{Scharnhorst}. Nel giro di poche ore le corazzate inglesi riuscirono a convergere verso la posizione dell'incrociatore che, invece di riavvicinarsi agli altri mezzi tedeschi, continuava a cercare il convoglio verso nord e portarono l'attacco che sarebbe culminato con l'affondamento dell'incrociatore prima di sera.

Vogliamo infine ricordare che la scelta stessa della data per lo sbarco in Normandia fu parzialmente dettata da messaggi decifrati grazie al lavoro dei crittoanalisti di Bletchley Park. Sin dal Maggio 1943, infatti, era stata stabilita la necessit\`a di un'invasione terrestre dell'Europa Nord-Occidentale, da compiere nella primavera del 1944. Il commando delle operazioni alleate vagli\`o varie possibili localit\`a per lo sbarco delle truppe da assalto, ma la decisione doveva essere presa con grande cura perch\'e era necessario scegliere un luogo che permettesse di mantenere, e poi espandere, delle teste di ponte Alleate sulle coste, senza subire l'attacco dell'Esercito corazzato tedesco. La scelta cadde sulla Normandia, ma allo stesso tempo venne pianificata ed attuata una serie di manovre diversive (la cosiddetta \emph{Operazione Bodyguard}) che potessero indurre i tedeschi a pensare che l'invasione sarebbe avvenuta in un momento successivo ed in un luogo diverso, ad esempio al Pas de Calais, o nel nord della Norvegia, o anche nel sud della Francia. Il ruolo dell'intelligence \emph{Ultra} in questa circostanza fu enorme perch\'e fu in grado di confermare che i tedeschi si aspettavano effettivamente lo sbarco a Calais e che i ``depistaggi'' Alleati avevano avuto successo, permettendo quindi ad inglesi e americani di confermare la Normandia come luogo dello sbarco e di stabilire la data reale in cui effettuarlo.

\section{Overview della cifratura di Enigma}\label{sec:overviewEnigma}

\subsection{Le componenti meccaniche di Enigma ed il loro effetto sulla cifratura}\label{sec:componenti}
La cifratura dei messaggi da parte di una macchina Enigma \`e il risultato di una serie di diversi meccanismi che si sovrappongono l'uno all'altro. Il cuore del meccanismo \`e l'applicazione di una sostituzione monoalfabetica che cambia per ciascun messaggio che viene inviato. Tale sostituzione, in termini matematici, non \`e altro che l'applicazione di una permutazione delle lettere dell'alfabeto (ossia un elemento di $S_{26}$) che ``rimescola'' le lettere del messaggio rimpiazzandole con altre. Da notare che nei messaggi i numeri venivano scritti per esteso come parole, quindi si trattava effettivamente di messaggi utilizzanti solamente $26$ caratteri in totale.

La determinazione precisa della sostituzione che viene applicata ad ogni lettera del messaggio \`e il risultato combinato di diverse componenti contenute all'interno di Enigma: da un lato i settaggi giornalieri, che risultano comuni a tutti i messaggi cifrati in quella data; dall'altro la scelta di una chiave specifica del singolo messaggio (o scelta liberamente dal marconista, o ottenuta come risultato di una complessa procedura standard a partire da alcune ``radici'' comuni).

Per dare un'idea pi\`u precisa di come venisse determinata e poi applicata la sostituzione delle singole lettere, dobbiamo volgere lo sguardo all'interno delle unit\`a Enigma. Il primo elemento da considerare sono i tre rotori che occupano la parte centrale del meccanismo. Mentre l'operatore cifra il messaggio che deve essere trasmesso, a ciascuna lettera del messaggio corrisponde il passaggio di un segnale elettrico attraverso i tre rotori. Tale segnale parte dal tasto premuto dall'operatore, attraversa i tre rotori in serie, poi viene riflesso dal riflettore (posto a destra dei rotori) ed infine ripassa attraverso i tre rotori in ordine inverso prima di andare ad illuminare la lettera cifrata corrispondente alla lettera cifrante.

\n In realt\`a, prima del passaggio tra i rotori e prima di arrivare al pannello di output, la lettera cifrata viene anche scambiata dalla \emph{plug board} se si tratta di una delle $20$ lettere soggette a scambio quel giorno. 

A complicare la questione interviene il fatto che, oltre alla selezione dei tre rotori, ciascuno con la propria posizione iniziale (cambiata anch'essa ogni giorno), la sostituzione monoalfabetica applicata da ciascun rotore \`e anche soggetta ad uno \emph{shift} legato alla posizione su cui viene settato il \emph{Ringstellung} (un anello incastrato nel bordo del rotore, in grado di modificare il risultato della cifratura di ciascun singolo elemento).

\n Questa cifratura di partenza comunque veniva utilizzata solo per la prima lettera del messaggio: ad ogni carattere criptato infatti il rotore pi\`u a destra avanzava di una posizione cambiando il tipo di trasformazione monoalfabetica da applicare alla lettera successiva. Come in una serie di ingranaggi, poi, dopo un certo numero di scatti il rotore di destra fa scattare quello centrale che, a sua volta, far\`a scattare in seguito quello di sinistra\ldots per un totale di $(26 \cdot 26 \cdot 26)$ diverse trasformazioni monoalfabetiche!

A chiusura della descrizione della cifratura, bisogna menzionare il \emph{Grundstellung} e la chiave di tre lettere che era peculiare di ciascun messaggio. Il \emph{Grundstellung} costituiva la posizione iniziale dei tre rotori per la giornata e veniva utilizzato come segue: all'inizio della trasmissione venivano trasmesse le tre lettere della chiave specifica del messaggio utilizzando la macchina settata con i rotori nella posizione del \emph{Grundstellung}. Dopo questi primi caratteri, i rotori venivano posizionati nelle posizioni indicate dalla chiave specifica del messaggio e questi andavano poi a determinare la chiave di cifratura applicata al messaggio vero e proprio.

Da notare che, in un primo tempo (dai primi anni '30 sino all'inizio del conflitto mondiale), la chiave specifica del messaggio veniva anche ritrasmessa una seconda volta con i nuovi settaggi, come meccanismo di conferma della corretta trasmissione. In questo modo le prime sei lettere del messaggio cifrato sarebbero risultate una ripetizione delle tre lettere della chiave cifrata dapprima con i settaggi del \emph{Grundstellung}, e poi con i rotori posizionati come prescritto dalla chiave del messaggio stesso. Con il proseguire della guerra tale abitudine venne abbandonata, rendendo inutili eventuali attacchi crittoanalitici basati su tale ripetizione.

Quando il messaggio veniva ricevuto, la procedura per recuperare il contenuto originale della comunicazione era speculare: la macchina Enigma veniva settata sulla configurazione del giorno, ossia secondo il \emph{Grundstellung}; le prime tre lettere venivano digitate nella macchina, ottenendo le corrispondenti lettere decifrate ossia la chiave del messaggio; poi i rotori venivano riposizionati come prescritto da queste tre lettere ed il resto del messaggio veniva inserito nella macchina recuperando l'originale. Si trattava quindi di un esempio di cifratura simmetrica, bench\'e ben pi\`u complesso del cifrario di Cesare o di quello di Vigen\`ere, descritti nel paragrafo~\ref{sec:antique}: volendo quantificare il numero di possibili chiavi che si potevano ottenere modificando le varie impostazioni iniziali della macchina si devono considerare le possibili scelte dei rotori e del loro ordine (un totale di $60$ combinazioni), le possibili posizioni iniziali dei rotori (un totale di $26^3$), i settaggi degli shift dovuti agli anelli (nuovamente $26^3$) ed il numero di combinazioni dei cavi scambiatori (pari a
$$
{26! \over 6! \cdot 10! \cdot 2^{10}} \,\simeq\, 1,5 \cdot 10^{14}
$$
in tutto), per un totale di circa $2,8 \cdot 10^{24}$ combinazioni.

In questo scritto ci concentriamo sull'analisi della procedura di identificazione di rotori e chiavi da parte degli analisti inglesi, piuttosto che sull'analisi legata a \emph{Ringstellung} e \emph{Steckerverbindungen}. Infatti dallo studio dei primi si coglie meglio l'impatto del teorema di Bayes, ed \`e pi\`u facile proporre alcuni esempi espliciti che sicuramente aiutano a comprendere il funzionamento della macchina.

\n Poniamo l'attenzione inoltre sul caso semplificato delle macchine Enigma con tre rotori, riservandoci di commentare il caso generale alla fine del nostro contributo.

\subsection{Un esempio semplificato}\label{sec:simply}
\begin{figure}[!htb]
\centering
\includegraphics[width=0.55\textwidth]{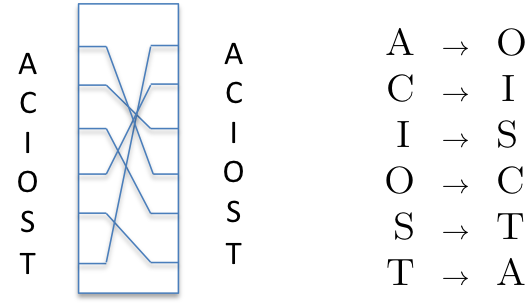}
\caption{Un rotore semplificato, agente su un alfabeto di 6 lettere.}
\label{fig:simply_rot}
\end{figure}

Per capire meglio il funzionamento della cifratura e dei tre rotori, semplifichiamo ulteriormente la situazione: supponiamo che il nostro alfabeto sia composto di sole 6 lettere \{\cypher{A}, \cypher{C}, \cypher{I}, \cypher{O}, \cypher{S}, \cypher{T}\} e andiamo ad analizzare il processo di crittazione. Abbiamo detto che ciascun rotore offre una \emph{sostituzione monoalfabetica}, ossia una cifratura in cui ciascuna lettera viene cifrata in un'altra, senza ripetizioni. Ad esempio, possiamo considerare il rotore mostrato in Figura~\ref{fig:simply_rot}, in cui sono mostrate esplicitamente le connessioni all'interno del rotore e le sostituzioni letterali risultanti.

Abbiamo poi detto che ad ogni lettera cifrata il rotore scatta di una posizione, producendo una nuova cifratura per la lettera successiva. 
\begin{figure}[!h]
\centering
\includegraphics[width=0.75\textwidth]{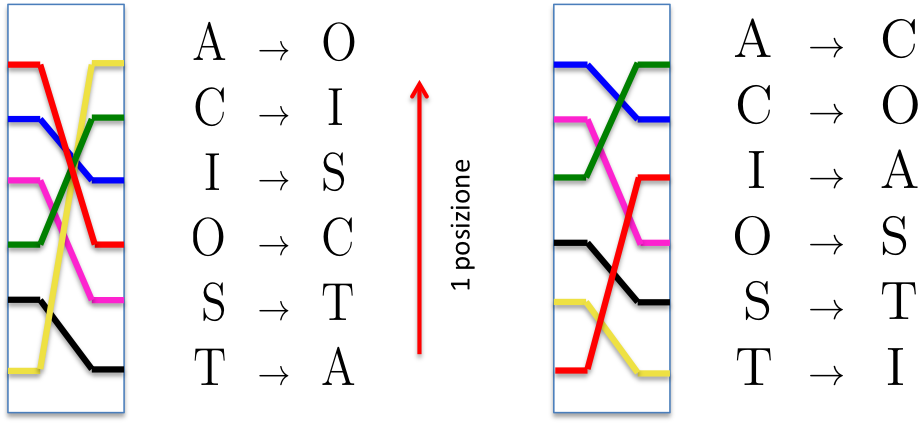}
\caption{La nuova trasformazione monoalfabetica dopo una rotazione.}
\label{fig:advance}
\end{figure}
Continuando ad analizzare il nostro rotore ``modello'', avremo dopo la cifratura della prima lettera un cambiamento come quello mostrato in Figura~\ref{fig:advance}, in cui viene anche mostrata esplicitamente la nuova cifratura ottenuta dopo l'avanzamento.
Nell'immagine abbiamo visualizzato le connessioni interne al rotore prima e dopo la rotazione (si dovrebbe immaginare il rettangolo in figura come se fosse avvolto a formare un cilindro, con le connessioni nella parte alta che risultano adiacenti a quelle nella parte bassa). Si nota ad esempio come la connessione nera, che inizialmente connette la penultima posizione in ``basso'' con l'ultima in ``basso'', finisca per collegare dopo lo scatto la terz'ultima posizione dal ``basso'' con la penultima dal ``basso''. Uno spostamento analogo viene applicato a ciascuna delle altre connessioni.

\begin{figure}[!htb]
\centering
\includegraphics[width=0.55\textwidth]{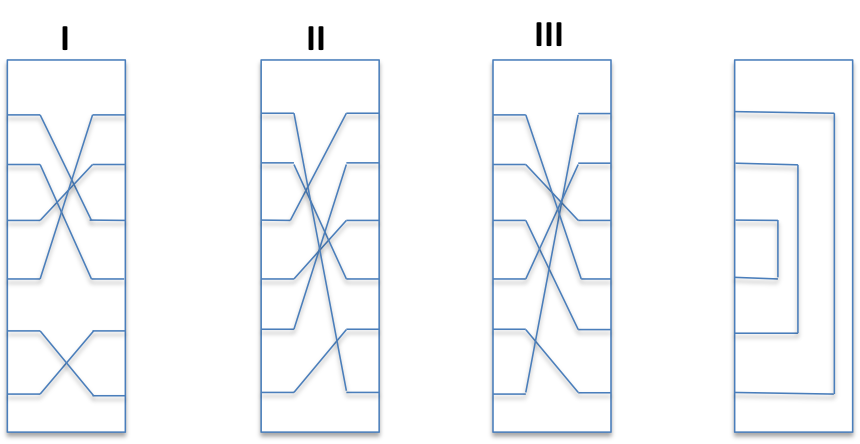}
\caption{Una macchina cifrante semplificata, agente su un alfabeto di 6 lettere.}
\label{fig:simply_complete}
\end{figure}

Volendo riprodurre il funzionamento della macchina Enigma con il nostro alfabeto ridotto, combiniamo tre rotori come il precedente, collocati nelle posizioni I, II e III, con un riflettore, ottenendo una configurazione come quella mostrata in Figura~\ref{fig:simply_complete}.

\n Supponiamo che nel nostro esempio si abbia uno scatto del rotore II solo dopo un giro completo del rotore III, ed uno scatto del rotore I solo dopo un giro completo del rotore II, per semplicit\`a. 

\begin{figure}[!htb]
\centering
\includegraphics[width=0.75\textwidth]{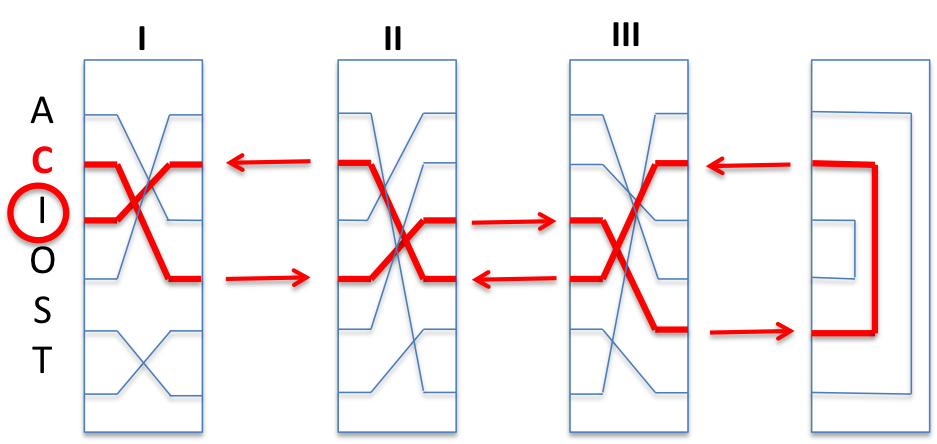}
\caption{La cifratura della prima lettera \cypher{C}.}
\label{fig:cypher1}
\end{figure}

Con questa macchina Enigma semplificate, proviamo a cifrare la sequenza \msg{CCC}. La cifratura della prima lettera \cypher{C} procederebbe come mostrato in Figura~\ref{fig:cypher1}, ottenendo una \cypher{I} come lettera cifrata, seguita da uno scatto del terzo rotore. 

\begin{figure}[!htb]
\centering
\includegraphics[width=0.75\textwidth]{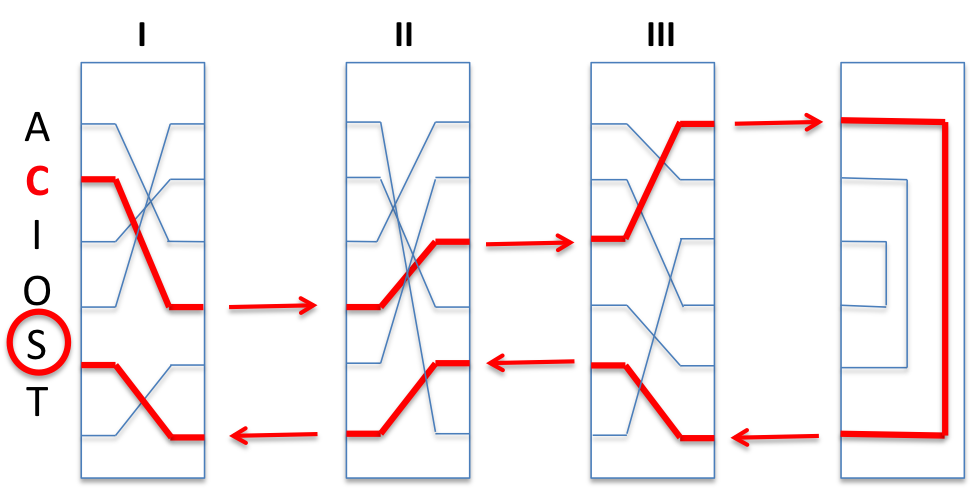}
\caption{La cifratura della seconda lettera \cypher{C}.}
\label{fig:cypher2}
\end{figure}

\n Come detto pi\`u volte in precedenza, tale scatto provoca un cambio della cifratura globale, cos\`\i{} che la seconda lettera \cypher{C} verr\`a cifrata in maniera differente rispetto alla prima. Mostriamo questa nuova situazione nella Figura~\ref{fig:cypher2}, da cui si nota come la nuova lettera \cypher{C} risulti cifrata in una lettera \cypher{S}, a cui segue un nuovo scatto del terzo rotore.
 
\begin{figure}[!htb]
\centering
\includegraphics[width=0.75\textwidth]{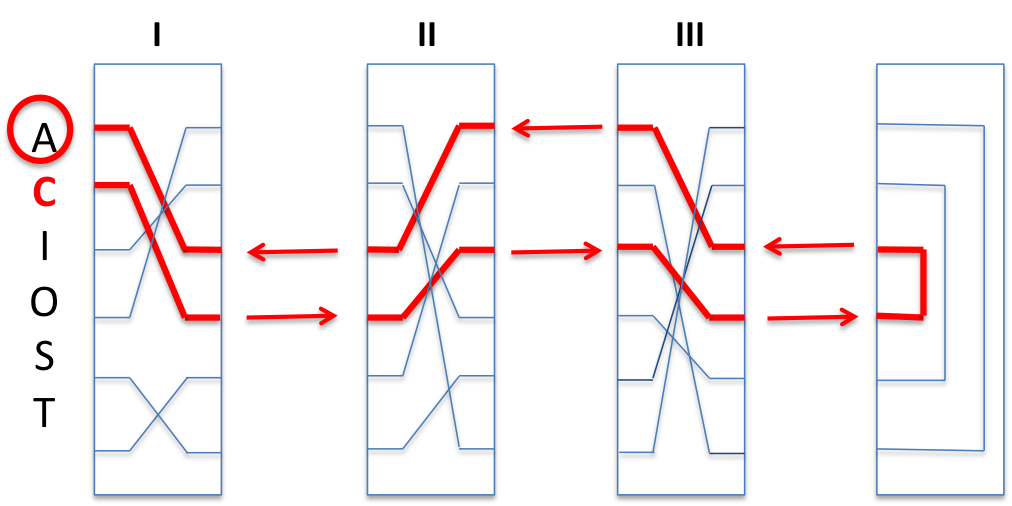}
\caption{La cifratura della terza lettera \cypher{C}.}
\label{fig:cypher3}
\end{figure}

\n Quando si procede a cifrare la terza lettera \cypher{C} otterremo il risultato illustrato in Figura~\ref{fig:cypher3}, ossia che
la lettera cifrata \`e questa volta una \cypher{A}.  
In conclusione la cifratura della sequenza \msg{CCC} risulta in \msg{ISA}.

A questo punto, se resettassimo i rotori nella posizione originale e cifrassimo la sequenza di lettere \msg{ISA}, vedremmo che il messaggio risultante sarebbe la sequenza originale \msg{CCC}. Questo perch\'e il ``doppio passaggio'' delle lettere attraverso i rotori e la presenza di un riflettore che non manda mai una lettera in se stessa garantiscono che partendo dai medesimi settaggi iniziali le operazioni di cifratura e decifratura siano sostanzialmente simmetriche. Quindi inserendo il messaggio in chiaro si otteniene il messaggio cifrato; inserendo il messaggio cifrato si riotteneva quello in chiaro (nel nostro esempio di alfabeto ridotto \`e sufficiente ripercorrere le linee nelle tre immagini invertendo le frecce rosse). 

Se da un lato tale simmetria semplifica il procedimento di trasmissione degli ordini strategici, dall'altro, come tutte le regolarit\`a nei meccanismi di cifratura, costituisce una debolezza della macchina, che potrebbero essere sfruttare da chi volesse invertire il processo. Questa debolezza era ovviamente nota ai tedeschi, ma essi ritennero che l'elevato numero di possibili chiavi di cifratura fosse sufficiente a garantire da solo l'inviolabilit\`a della cifratura.

Ricapitolando, abbiamo visto che a partire dalla struttura meccanica della macchina Enigma, si possono ricavare alcune propriet\`a della cifratura:
\begin{itemize}
\item ogni rotore applica uno \emph{scambio monoalfabetico};
\item dopo lo scambio, c'\`e uno scatto di uno o pi\`u rotori (che modifica la cifratura della prossima lettera del messaggio);
\item la cifratura non si pu\`o ripetere in messaggi pi\`u corti di $26^3$ lettere ($\sim 17600$);
\item valgono le seguenti propriet\`a:
\begin{enumerate}
\item nessuna lettera pu\`o essere cifrata in se stessa [\emph{non--identit\`a}];
\item se $L_1$ \`e cifrata in $L_2$, allora anche $L_2$ \`e cifrata in $L_1$ [\emph{simmetria}].
\end{enumerate}
\end{itemize}

\subsection{I cavi scambiatori}\label{sec:cables}
In aggiunta alla cifratura polialfabetica applicata dai tre rotori e dal riflettore presenti al suo interno, la macchina Enigma permetteva di scambiare un certo numero di coppie di lettere configurando opportunamente i cavi del pannello frontale della macchina (\emph{plug board}). Nella versione originale di Enigma, quella decifrata dai polacchi nei primi anni '30, le coppie di lettere scambiate erano solo sei, mentre nella versione utilizzata dai tedeschi durante la Seconda Guerra Mondiale esse erano dieci. 
Per come era collegata la \emph{plug board} ai restanti meccanismi lo scambio di lettere veniva applicato due volte: una prima volta sulla lettera in chiaro del messaggio (ossia prima dell'azione della cifratura dei rotori) ed una seconda volta sulla lettera cifrata del messaggio (ossia dopo l'azione della cifratura dei rotori).

Per capire meglio il modo in cui i cavi scambiatori intervenivano, andiamo nuovamente a considerare il nostro alfabeto ridotto del paragrafo~\ref{sec:simply}, ossia supponiamo di utilizzare una macchina Enigma grandemente semplificata in cui i messaggi possono solo contenere le lettere \{\cypher{A}, \cypher{C}, \cypher{I}, \cypher{O}, \cypher{S}, \cypher{T}\}. Supponiamo che ci possano essere due coppie di lettere scambiate, in questa macchina semplificata, e supponiamo che la scelta cada sulle coppie di lettere \{\cypher{A}, \cypher{C}\} e \{\cypher{S}, \cypher{O}\}. Andiamo a testare la nostra macchina cifrando la sequenza \msg{AAA} a partire dalla medesima configurazione iniziale del paragrafo~\ref{sec:simply}, ossia quella rappresentata nella Figura~\ref{fig:simply_cabled}, in cui abbiamo aggiunto l'azione dello scambio di lettere nella colonna sinistra. Otteniamo che:
 
\begin{figure}[!htb]
\centering
\includegraphics[width=0.75\textwidth]{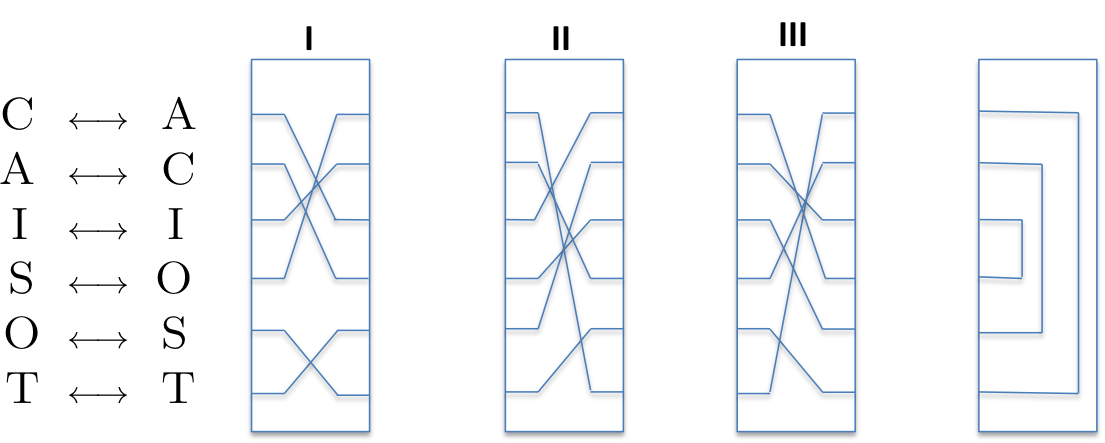}
\caption{La macchina cifrante semplificata a cui abbiamo aggiunto due cavi scambiatori.}
\label{fig:simply_cabled}
\end{figure}
 
\begin{itemize}
\item La prima \cypher{A} verr\`a scambiata con \cypher{C} per effetto della \emph{plug board}, tale \cypher{C} verr\`a cifrata in una \cypher{I} come nel paragrafo precedente, e la \cypher{I} verr\`a lasciata immutata dalla \emph{plug board}.
\item La seconda \cypher{A} verr\`a scambiata con \cypher{C} per effetto della \emph{plug board}, tale \cypher{C} verr\`a cifrata in una \cypher{S} come nel paragrafo precedente (perch\'e il terzo rotore \`e avanzato di una posizione), e la \cypher{S} verr\`a cambiata in \cypher{O} dalla \emph{plug board}.
\item La terza \cypher{A} verr\`a scambiata con \cypher{C} per effetto della \emph{plug board}, tale \cypher{C} verr\`a cifrata in una \cypher{A} come nel paragrafo precedente (perch\'e il terzo rotore \`e avanzato di un'ulteriore posizione), e la \cypher{A} verr\`a cambiata in \cypher{C} dalla \emph{plug board}.
\end{itemize}
Quindi il risultato della cifratura della sequenza \msg{AAA} sar\`a la sequenza \msg{IOC}.

Osserviamo che le propriet\`a di \emph{non-identit\`a} e di \emph{simmetria} delineate nel paragrafo~\ref{sec:simply} rimangono altrettanto valide anche se si applicano scambi di coppie di lettere tramite la \emph{plug board}. Questo accade perch\'e viene applicato un duplice scambio, sia sul messaggio in chiaro che sul messaggio cifrato.

\subsection{I settaggi giornalieri}\label{sec:settings}
Ogni giorno, l'operatore incaricato di inviare i messaggi doveva:
\begin{itemize}
\item prendere il foglio con i settaggi giornalieri della macchina;
\item scegliere i rotori prescritti e posizionarli nella sequenza corretta;
\item posizionare i \emph{Ringstellung} come prescritto;
\item collegare i cavi della \emph{plug board} secondo gli accoppiamenti segnati;
\item annotare il \emph{Grundstellung} del giorno (la procedura era pi\`u complessa per la \emph{Kriegsmarine}).
\item scegliere una chiave specifica per il messaggio (3 lettere).
\end{itemize}
Quindi per ogni messaggio c'erano {\bf due} chiavi da tre lettere:
\begin{itemize}
\item il \emph{Grundstellung} che era la chiave comune a {\bf tutti} i messaggi del giorno;
\item un secondo gruppo di tre lettere specifico per ciascun messaggio e scelto dall'operatore.
\end{itemize}
Come gi\`a menzionato, entrambe le chiavi rappresentavano una posizione iniziale per ciascuno dei 3 rotori di Enigma.

A questo punto l'operatore era pronto a trasmettere. La procedura era quindi la seguente, dal suo 
punto di vista:
\begin{enumerate}
\item settare la posizione iniziale dei rotori in base al \emph{Grundstellung};
\item cifrare la chiave specifica del messaggio con tali settaggi (3 lettere);
\item settare i rotori alla posizione della chiave specifica;
\item cifrare il messaggio vero e proprio (fino al 1942 circa, il messaggio era preceduto da una ripetizione della chiave, stavolta cifrata con i nuovi settaggi).
\end{enumerate}

\`E interessante notare come, per i messaggi dell'Esercito e dell'Aviazione tedeschi, l'arbitrariet\`a nella scelta della chiave del messaggio si rivel\`o in alcuni casi una delle poche debolezze del sistema: essendo lasciata la scelta a colui che doveva trasmettere il messaggio, capitava che questi utilizzasse talvolta sequenze di senso compiuto (pi\`u semplici da ricordare) o nomi di persona, facilitando il compito dei decifratori qualora essi avessero riconosciuto la chiave. Ad esempio, si scopr\`\i{} che uno degli operatori in Africa era solito utilizzare la medesima chiave per settimane, nonostante gli ordini dei suoi superiori, e una volta decifrata quella, divenne praticamente immediato decifrare tutte le comunicazioni inviate da quello specifico operatore.

\section{Overview del lavoro di decifratura di Enigma}\label{sec:decifratura}

Rejewski e gli altri decifratori polacchi avevano intuito che il problema di decifrare Enigma si poteva ``fattorizzare'' in problemi separati ed \emph{indipendenti}:
\begin{itemize}
\item capire l'ordine dei tre rotori;
\item identificare le lettere scambiate tramite la \emph{plug board} (all'epoca 6 coppie);
\item identificare la chiave giornaliera.
\end{itemize}
In particolare l'ordine dei rotori poteva essere dedotto da considerazioni non legate al contenuto del messaggio, ma ottenibili dal solo confronto di (numerosi) messaggi cifrati con chiavi simili, sfruttando 
la presenza della chiave ad inizio messaggio.

\n Gli inglesi si trovarono a dover fare i conti con l'aggiunta (dal 1939) di altri due rotori tra cui venivano scelti i tre usati e altri 4 cavi di scambio nella \emph{plug board}, che portavano da 6 a 10 le coppie di lettere scambiate dalla macchina. L'aumento delle configurazioni possibili rendeva necessario un nuovo metodo di riduzione delle possibilit\`a da testare in quanto altrimenti non era possibile isolare un'unica chiave di decifratura valida (quella corretta) prima che i tedeschi la modificassero.

I matematici di Bletchley Park (capitanati da Turing e Good) adottarono varie strategie per attaccare il problema dell'identificazione dei rotori e del loro ordine. In questa sezione vogliamo analizzare in particolare: 
\begin{enumerate}
\item \emph{Banburismus};
\item \emph{Scritchmus}.
\end{enumerate}
Lo scopo di tali strategie era quello di identificare quale fosse il terzo rotore utilizzato in quella specifica giornata, tra quelli a disposizione. Talvolta anche il secondo rotore poteva essere identificato applicando una versione modificata delle medesime idee. 
Naturalmente l'identificazione di uno (o due) dei rotori utilizzati significava una sensibile riduzione nel numero di combinazioni che le macchine \emph{bomba} dovevano testare per trovare l'effettiva chiave di cifratura.

Delle due tecniche menzionate, la prima sar\`a quella a cui dedicheremo la maggiore attenzione, cercando di analizzarne la maggior parte dei dettagli nel paragrafo~\ref{sec:BayesBanburismus}, perch\'e \`e quella in cui la regola di Bayes e l'approccio probabilistico giocano un ruolo chiave. Cercheremo tuttavia di descrivere nei paragrafi~\ref{sec:scritchmus}--\ref{sec:generale} anche un esempio di decifratura ``completo'', in cui siano dati opportuni dettagli della seconda tecnica e del modo in cui tali tecniche venivano sfuttate dalle macchine \emph{bomba}, per offrire un pi\`u preciso quadro d'insieme della procedura.

\subsection{\emph{Banburismus}}\label{sec:banburismus}
Avendo a disposizione numerosi messaggi cifrati ogni giorno (intercettati dall'intelligence), l'idea alla base di questa tecnica \`e quella di sfruttarli per identificare il terzo rotore e, se possibile, anche il secondo tra quelli disponibili. Lo spunto di partenza \`e quello di usare messaggi che siano stati cifrati con chiavi ``simili''.

Supponiamo ad esempio di avere due messaggi in cui le prime due lettere cifrate sono le medesime e la terza no:
$$
\begin{array}{c}
\mbox{\cypher{PMG    TZUYJZSLAPEIXM\ldots}}\\
\mbox{\cypher{PMQ    KVRPOLXLYTWSIL\ldots}}
\end{array}
$$
Poich\'e il \emph{Grundstellung} giornaliero \`e lo stesso, le due sequenze \cypher{PMG} e \cypher{PMQ} sono sicuramente state cifrate a partire dalla stessa disposizione dei rotori. Il fatto che le prime due lettere siano uguali significa quindi che le prime due lettere della chiave del messaggio sono anch'esse uguali! Solo la terza lettera della chiave del messaggio \`e diversa (perch\'e essa viene cifrata in un caso con \cypher{G} e nell'altro con \cypher{Q}).

\n Grazie alle considerazioni fatte nei paragrafi~\ref{sec:componenti} e~\ref{sec:simply}, sulla struttura meccanica delle macchine Enigma, possiamo ora dedurre che ci\`o che rende diverse le cifrature applicate ai due messaggi \`e solo la terza lettera diversa nelle sequenze \cypher{PMG} e \cypher{PMQ}. Se noi potessimo in qualche modo ``riavvolgere'' il terzo rotore cos\`\i{} da avere che anche la terza lettera della chiave fosse la stessa, allora otterremmo la medesima cifratura in entrambi i messaggi. Avremmo quindi due messaggi che si direbbero \emph{in depth} (= cifrati con gli stessi settaggi), la cui cifratura procederebbe di pari passo almeno fino al momento in cui anche il secondo rotore avanza. Questa osservazione, a prima vista banale, si rivel\`o in realt\`a cruciale: si riducevano le differenze di cifratura tra i due messaggi ad una semplice ``distanza'' tra le terze lettere della chiave e si dava un suggerimento su come confrontare i due messaggi cifrati.

L'idea alla base del \emph{Banburismus} consisteva quindi nel provare a sovrapporre i due messaggi, facendoli scorrere l'uno sull'altro come se volessimo riavvolgere il terzo rotore ignorando un certo numero di lettere cifrate ad inizio messaggio, e valutare se il risultato della parte comune potesse essere o meno \emph{in depth} a partire dal numero di lettere uguali che si trovavano in posizioni uguali. 

\begin{figure}[!htb]
\centering
\includegraphics[width=0.55\textwidth]{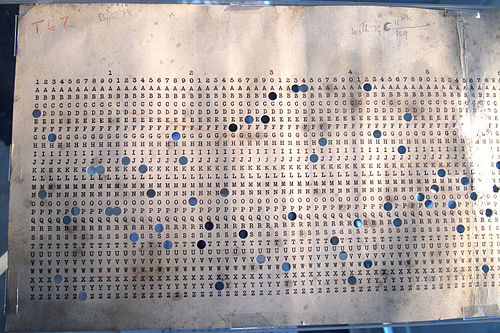}
\caption{Un \emph{Banbury sheet} con ``marcato'' il messaggio cifrato. (fonte: \emph{Wikipedia}, autore: \emph{TedColes}, licenza: \href{https://creativecommons.org/licenses/by-sa/4.0/}{CC BY-SA-4.0})}
\label{fig:banbury}
\end{figure}

A ciascuna posizione di un messaggio rispetto all'altro, rappresentabile come una distanza tra le terze lettere del messaggio (nel nostro caso una distanza tra \cypher{G} e \cypher{Q}, come ad esempio $G=Q+1$, $G=Q+2$, ecc.), si ottiene una possibilit\`a che essi siano effettivamente \emph{in depth}. Non avendo per\`o un modo ``certo'' per capire quando i messaggi in effetti lo sono, il meglio che si possa fare \`e di fornire una valutazione di quanto sia probabile che ciascuna posizione corrisponda ad un ``allineamento'' del terzo rotore sulla base del numero di lettere coincidenti trovate nei due messaggi in quella posizione. E lo strumento naturale per quantificare la probabilit\`a in questo contesto \`e il teorema di Bayes: a ciascuna posizione \`e possibile associare il {\bf fattore di Bayes-Turing} o {\bf peso delle evidenze}, che \`e legato alla \emph{odd} (o \emph{quota}) che si pu\`o associare alla verosimiglianza che i due messaggi siano \emph{in depth}, come vedremo nel paragrafo~\ref{sec:BayesBanburismus}. Dal confronto dei fattori di Bayes--Turing associati a ciascuna posizione \`e possibile ottenere un ranking tra le posizioni e, a loro volta, i risultati del ranking permettono poi di restringere effettivamente, nella fase di \emph{Scritchmus}, il numero di possibili scelte per il terzo rotore.

\`E interessante osservare che la comparazione dei diversi messaggi veniva effettuata fisicamente dalla sovrapposizione di fogli come quelli mostrati in Figura~\ref{fig:banbury}.
Ciascun messaggio intercettato veniva rappresentato su un foglio simile a quello mostrato, forando lettera per lettera la striscia di carta. Una volta completata questa ``trascrizione'', membri del personale ausiliario della struttura procedevano manualmente ai confronti tra le possibili posizioni relative di ciascuna coppia di messaggi, facendo scorrere un foglio sopra all'altro e osservando quante coppie di buchi risultassero sovrapposte nelle varie posizioni, corrispondenti alla sovrapposizione di lettere uguali in posizioni uguali. Come vedremo nella sezione~\ref{sec:FocusBayes}, da tale numero di coincidenze si otteneva, con la regola di Bayes, lo \emph{score} corrispondente alla posizione.

Chiudiamo il paragrafo con un aneddoto: il nome \emph{Banburismus} scelto per indicare tale tecnica proviene dal nome della cittadina di Banbury, poco distante da Bletchley Park, in cui venivano prodotte le striscie di carta utilizzate per comparare i diversi messaggi. 

\subsection{\emph{Scritchmus}}\label{sec:scritchmus}
Il risultato della fase precedente si pu\`o sintetizzare in una tabella del tipo:

\begin{center}
\begin{tabular}{|l|l|}
\hline
Shift	& Fattore B-T\\
\hline
\hline
$G=K+4$ & \odds{600:1} on\\
$B=N-24$ & \odds{500:1} on\\
$M=Q+16$ & \odds{200:1} on\\
$C=Q+18$ & \odds{22:1} on\\
$J=L+24$ & \odds{9:1} on\\
$C=M+2$ & \odds{7:1} on\\
$G=V+9$ & \odds{9:2} on\\
$A=Z-10$ & \odds{4:1} on\\
\ldots  & \ldots\\
\hline
\end{tabular}
\end{center}

In tale tabella viene rappresentato il ``peso probabilistico'', o \emph{fattore di Bayes-Turing} che avevamo gi\`a menzionato nel paragrafo precedente, corrispondente alla pi\`u probabile posizione \emph{in depth} per ciascuna coppia di messaggi aventi le medesime due prime lettere della chiave. Per farsi un'idea di quanto affidabile possa essere tale score, si pensi che il medesimo fattore viene utilizzato nei pi\`u moderni esperimenti di fisica per analizzare i dati raccolti e che valori superiori a \odds{50:1} vengono considerati evidenze ``molto forti'' e valori superiori a \odds{100:1} vengono considerati evidenze ``decisive''. Tipicamente nel contesto dei messaggi di Enigma un fattore di \odds{50:1} o superiore veniva trattato come una ``quasi certezza''.

La seconda fase della procedura seguita a Bletchley Park, lo \emph{Scritchmus}, consisteva a questo punto nell'utilizzare la tabella per costruire delle ``catene'' di lettere che possano portare all'identificazione dell'alfabeto cifrante corrispondente ai settaggi di Enigma considerati. Questo, a sua volta, permetter\`a di individuare quale rotore \`e stato inserito come terzo nella giornata considerata. Vogliamo fornire un esempio guida della costruzione di queste catene usando la tabella di esempio mostrata sopra. Si parte dalle righe della tabella con lettere comuni e si costruisce una sequenza in cui le lettere distino esattamente tanti spazi quanta era la distanza pi\`u probabile individuata nella fase del \emph{Banburismus}. Quindi dalle righe $G=K+4$ e $G=V+9$ si pu\`o costruire la catena:
$$
\begin{array}{c}
\mbox{\cypher{V - - - - K - - - G}}
\end{array}
$$
in cui abbiamo posizionato la \cypher{G} ad una distanza di 4 posizioni dalla \cypher{K} e la \cypher{V} ad una distanza di 9 posizioni dalla \cypher{G}. Analogamente dalle righe $M=Q+16$ e $C=Q+18$ si pu\`o costruire la catena:
$$
\begin{array}{c}
\mbox{\cypher{M - C - - - - - - - Q}}
\end{array}
$$
Si devono poi posizionare tali sequenze all'interno di un possibile alfabeto cifrante che sia compatibile con le regole di \emph{simmetria} e \emph{non-identit\`a} che contraddistinguono la cifratura di Enigma (le propriet\`a individuate nel paragrafo~\ref{sec:simply}). Delle $26$ posizioni possibili in cui possiamo piazzare ciascuna catena (una per ciascuna lettera a cui possiamo far corrispondere le lettere \cypher{V} ed \cypher{M}, rispettivamente), infatti, solo alcune sono compatibili con tali regole. Consideriamo ad esempio il seguente posizionamento:
$$
\begin{array}{l}
\mbox{\cypher{A B C D E F G H I J K L M N O P Q R S T U V W X Y Z}}\\
\mbox{\cypher{\phantom{A B C D E }V - - - - K - - - G}}
\end{array}
$$
Esso non \`e valido perch\'e trasformerebbe la lettera \cypher{K} in se stessa. Similmente non \`e valido il posizionamento mostrato di seguito:
$$
\begin{array}{l}
\mbox{\cypher{A B C D E F G H I J K L M N O P Q R S T U V W X Y Z}}\\
\mbox{\cypher{\phantom{A B C D E F }V - - - - K - - - G}}
\end{array}
$$
poich\'e l'alfabeto risultante trasformerebbe la \cypher{G} in \cypher{V} e poi la \cypher{P} in \cypher{G}, violando la simmetria di qualunque cifratura generata con Enigma.

Vari altri posizionamenti sono invece compatibili con queste propriet\`a e li consideriamo tutti equalmente validi. Ad esempio, non ci sono contraddizioni nei posizionamenti:
$$
\begin{array}{l}
\mbox{\cypher{A B C D E F G H I J K L M N O P Q R S T U V W X Y Z}}\\
\mbox{\cypher{\phantom{A }V - - - - K - - - G}}
\end{array}
$$
e
$$
\begin{array}{l}
\mbox{\cypher{A B C D E F G H I J K L M N O P Q R S T U V W X Y Z}}\\
\mbox{\cypher{- G \phantom{D E F G H I J K L M N O P Q R S }V - - - - K - -}}
\end{array}
$$
Sfruttando la simmetria possiamo anche completare nel primo caso lo scambio di \cypher{V} in \cypher{B}, e nel secondo \cypher{V} in \cypher{S}, \cypher{K} in \cypher{X} e \cypher{G} in \cypher{B}, ottenendo:
$$
\begin{array}{l}
\mbox{\cypher{A B C D E F G H I J K L M N O P Q R S T U V W X Y Z}}\\
\mbox{\cypher{- V - - - - K - - - G - - - - - - - - - - B - - - -}}
\end{array}
$$
e
$$
\begin{array}{l}
\mbox{\cypher{A B C D E F G H I J K L M N O P Q R S T U V W X Y Z}}\\
\mbox{\cypher{- G - - - - B - - - X - - - - - - - V - - S - K - -}}
\end{array}
$$
Per cercare di completare l'alfabeto cifrante, cerchiamo ora di posizionare la seconda catena e, come fatto con la prima, scartiamo posizionamenti che non siano compatibili con la cifratura. Ad es.:
$$
\begin{array}{l}
\mbox{\cypher{A B C D E F G H I J K L M N O P Q R S T U V W X Y Z}}\\
\mbox{\cypher{- G - - - - B - - - X - - - - - - - V - - S - K - -}}\\
\mbox{\cypher{\phantom{A B }M - C - - - - - - - Q}}
\end{array}
$$
va scartata perch\'e \cypher{C} in \cypher{M} non \`e compatibile con \cypher{E} in \cypher{C}. D'altra parte \`e invece valido il posizionamento:
$$
\begin{array}{l}
\mbox{\cypher{A B C D E F G H I J K L M N O P Q R S T U V W X Y Z}}\\
\mbox{\cypher{- G - - - - B - - - X - - - - - - - V - - S - K - -}}\\
\mbox{\cypher{\phantom{A B C }M - C - - - - - - - Q}}
\end{array}
$$
che permette di arrivare a:
$$
\begin{array}{l}
\mbox{\cypher{A B C D E F G H I J K L M N O P Q R S T U V W X Y Z}}\\
\mbox{\cypher{- G F M - C B - - - X - D Q - - N - V - - S - K - -}}
\end{array}
$$
e cos\`\i{} via continuando a scorrere la tabella degli shift alla ricerca di nuove catene. 

Al termine della procedura possono esserci varie situazioni:
\begin{itemize}
\item Nessun alfabeto cifrante \`e compatibile con le catene usate: in questo caso significa che
una o pi\`u delle nostre valutazioni \`e errata e dobbiamo ripetere la fase di \emph{Scritchmus} scartando uno degli \emph{shift}, partendo naturalmente da quelli che hanno \emph{peso} pi\`u basso.
\item Un solo alfabeto cifrante \`e compatibile con le catene, ad es.
$$
\begin{array}{l}
\mbox{\cypher{A B C D E F G H I J K L M N O P Q R S T U V W X Y Z}}\\
\mbox{\cypher{R G F M J C B T U E X Z D Q W Y N A V H I S O K P L}}
\end{array}
$$
\item Pi\`u alfabeti sono compatibili con le catene.
\end{itemize}
Nel secondo e nel terzo caso, possiamo allora passare ad analizzare quale (o quali) rotori siano compatibili con l'alfabeto trovato. Era infatti noto che ciascuno dei rotori usati fino al 1942 erano caratterizzati da differenti ``punti di scatto'' in cui il rotore forza il suo omologo a sinistra ad avanzare di una posizione. Tale conoscenza non \`e di immediata utilit\`a nel decifrare i messaggi perch\'e le lettere corrispondenti agli effettivi scatti potevano essere modificate dalla posizione dei \emph{Ringstellung} del primo e secondo rotore, ma pu\`o essere utilizzata in questo contesto perch\'e se un punto di scatto fosse caduto all'interno di una delle catene utilizzate per costruire l'alfabeto cifrante, da quel punto in poi i messaggi non sarebbero pi\`u stati \emph{in depth} (da quel punto in poi, infatti, solo una delle due macchine avrebbe avuto lo scatto del secondo rotore) e questo avrebbe in generale portato ad un peso inferiore come risultato del \emph{Banburismus}. Nel nostro \emph{toy example} avevamo utilizzato le seguenti catene (tra le altre possibili):
$$
\begin{array}{l}
\mbox{\cypher{\phantom{A B C D}|2\phantom{ F G H I J}|4\phantom{ L M N O P Q}|1\phantom{ S T U V}|3\phantom{ X Y Z}|5}}\\
\mbox{\cypher{A B C D E F G H I J K L M N O P Q R S T U V W X Y Z}}\\
\mbox{\cypher{R G F M J C B T U E X Z D Q W Y N A V H I S O K P L}}\\
\mbox{\cypher{- + \phantom{C D E F G H I J K L M N O P Q R }+ - - - - - - - }}\\
\mbox{\cypher{\phantom{A B C }+ - - - - - - - - - +}}\\
\mbox{\cypher{\phantom{A B C D }+ - - - - + }}
\end{array}
$$
In questo caso particolare si deduce che solo il rotore \texttt{1} \`e compatibile con le tre catene mostrate, perch\'e gli altri rotori hanno punti di scatto che ``interromperebbero'' la catena. 

Ancora in questa fase \`e possibile che non si trovino rotori compatibili con le catene: in tal caso significa che almeno una delle catene da noi utilizzate non era corretta e dobbiamo ripetere il processo di costruzione dell'alfabeto cifrante compatibile dopo aver scartato almeno una delle valutazioni usate per costruire le catene, partendo da quelle con peso pi\`u basso.

\subsection{Il rotore centrale, le \emph{Cryptologic Bomba} \& i \emph{crib}}\label{sec:crib}
L'approccio presentato nei paragrafi precedenti per identificare il rotore utilizzato come terzo (nella posizione destra), combinando la tecnica del \emph{Banburismus} con quella dello \emph{Scritchmus}, poteva poi essere adattata all'analisi di quale potesse essere il rotore centrale. Ripetendo infatti un ragionamento simile, attraverso il confronto di coppie di messaggi in cui era il secondo carattere della chiave di messaggio ad essere differente, si poteva spesso identificare anche quale fosse il rotore II del settaggio giornaliero, o per lo meno scartare alcuni dei rotori possibili.

Giunti a tale punto, toccava alle macchine \emph{bomba} mettersi all'opera per cercare l'effettiva chiave di decifrazione dei messaggi di ciascun giorno. Le macchine \emph{bomba} erano infatti capaci di selezionare la configurazione compatibile (o le poche configurazioni compatibili) sia con i rotori individuati che con i cosiddetti \emph{crib}: questi erano parti del messaggio cifrato di cui si poteva congetturare il significato sulla base della conoscenza dei messaggi militari tedeschi e che permettevano di scartare un gran numero di possibili configurazioni iniziali della macchina cifrante. In dettaglio, dall'analisi condotta tramite le macchine \emph{bomba} era possibile confermare come compatibile o scartare una particolare sequenza di rotori presenti, la loro posizione iniziale (\emph{Grundstellung}) e una specifica coppia di lettere scambiate tramite \emph{plug board}.

Un esempio di \emph{crib} era dato da espressioni quali \msg{Keine besonderen Ereignisse} (= niente da segnalare) o espressioni contenenti la parola \msg{wetter} (= condizioni meteo) che tipicamente gli operatori trasmettevano alla medesima ora ogni giorno, all'interno di messaggi che possedevano una struttura quasi sempre fissa: se si individuava la sua posizione in un messaggio, poi si poteva sfruttare anche nei giorni successivi! In questo caso, il grande numero di messaggi di routine inviati nei rapporti militari giornalieri al loro comando giocava contro la sicurezza dei messaggi stessi. Un altro esempio di \emph{crib} era dato dalla parola \msg{EINS} che compariva con alta frequenza nei dispacci, sia quando si doveva inviare una posizione geografica, sia quando si doveva utilizzarlo come quantificatore: gli alleati produssero un intero ``catalogo'' dei modi in cui la parola poteva venire cifrata dai vari rotori, e tale catalogo poteva dare conferma sull'effettivo utilizzo di un settaggio iniziale piuttosto che un altro.

L'effettiva modalit\`a di utilizzo dei \emph{crib}, strettamente correlata con il modo in cui Turing e gli altri crittoanalisti progettarono le loro migliorie alle macchine \emph{bomba} polacche, meriterebbe probabilmente un articolo a parte (rimandiamo il lettore maggiormente interessato a questo aspetto a~\cite{EnigmaSpec2,EnigmaSpec3}). In questa sede forniamo solo un breve esempio di utilizzo di tali informazioni per ridurre il numero di possibili coppie di lettere che venivano scambiate dal \emph{Steckerverbindungen}, preso da~\cite{EnigmaSpec2}. Immaginiamo quindi di poter congetturare che la frase \msg{WETTERVORHERSAGEBISKAYA} (= condizioni meteo di Biskaya) sia cifrata nella sequenza \msg{RWIVTYRESXBFOGKUHQBAISE}. \`E allora possibile costruire una tabella come la seguente:
$$
\begin{array}{l}
\mbox{\cypher{1\ \ 2\ \ 3\ \ 4\ \ 5\ \ 6\ \ 7\ \ 8\ \ 9\ 10\ 11\ 12\ 13\ 14\ 15\ 16\ 17\ 18\ 19\ 20\ 21\ 22\ 23}}\\
\mbox{\cypher{R\ \ W\ \ I\ \ V\ \ T\ \ Y\ \ R\ \ E\ \ S\ \ X\ \ B\ \ F\ \ O\ \ G\ \ K\ \ U\ \ H\ \ Q\ \ B\ \ A\ \ I\ \ S\ \ E}}\\
\mbox{\cypher{W\ \ E\ \ T\ \ T\ \ E\ \ R\ \ V\ \ O\ \ R\ \ H\ \ E\ \ R\ \ S\ \ A\ \ G\ \ E\ \ B\ \ I\ \ S\ \ K\ \ A\ \ Y\ \ A}}
\end{array}
$$

\begin{figure}[!htb]
\centering
\includegraphics[width=0.55\textwidth]{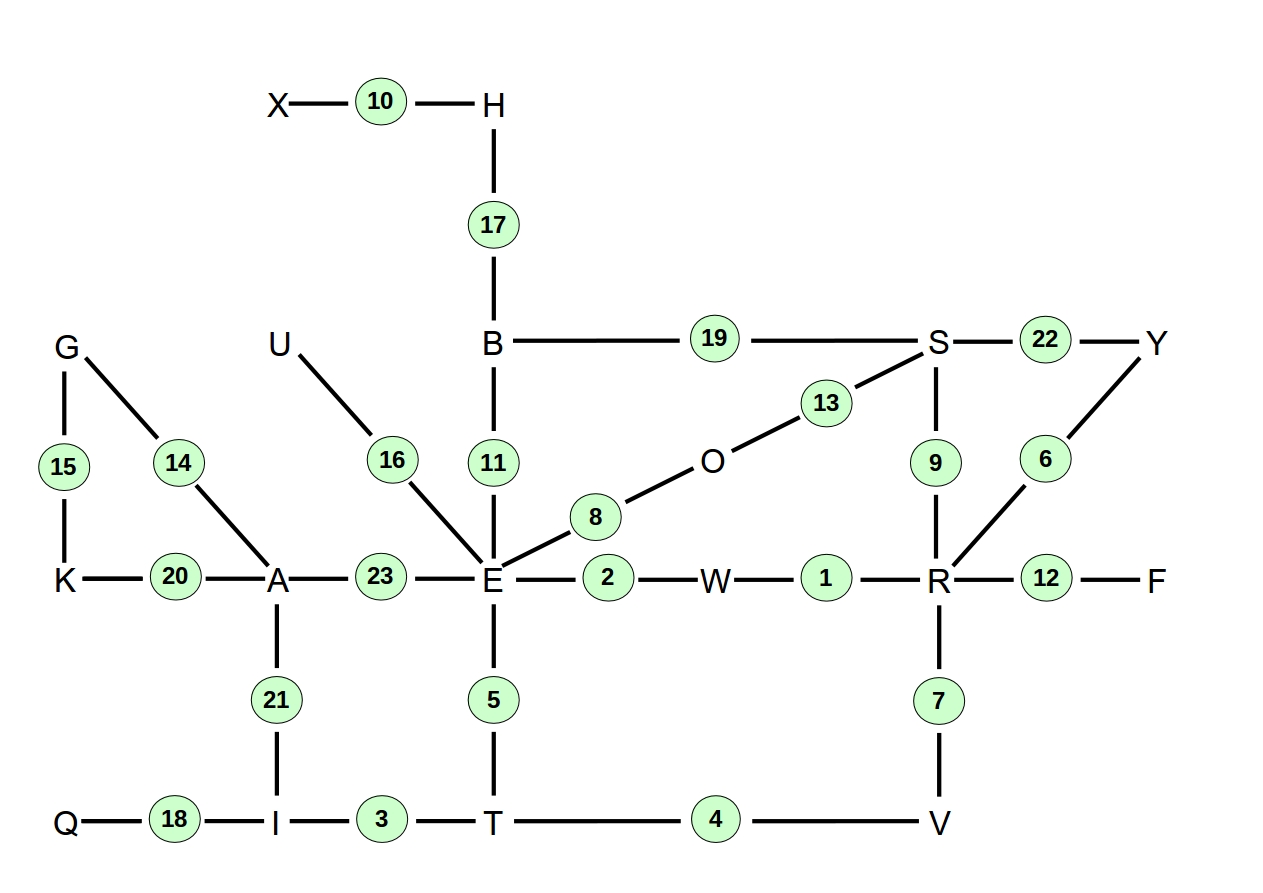}
\caption{Il grafo relativo al \emph{crib} ipotizzato.}
\label{fig:crib_graph}
\end{figure}

Andando a riportare le coppie di lettere trasformate in ciascuna posizione (la posizione \`e rilevante perch\'e dopo ciascuna lettera almeno uno dei rotori avanza di uno scatto e quindi ogni volta una cifratura differente verr\`a applicata alla lettera successiva), \`e possibile costruire una rappresentazione ``a grafo'' del \emph{crib} e della sua cifratura come quella in Figura~\ref{fig:crib_graph}.
 
Si osservi ad esempio che la lettera \cypher{T}, che compare nell'ultima riga del grafo, \`e connessa alle lettere \cypher{I} (alla terza posizione), \cypher{V} (alla quarta posizione) ed \cypher{E} (alla quinta posizione). Le relazioni sono bidirezionali perch\'e a posizione fissata vale la propriet\`a di \emph{simmetria} della cifratura.

In questo tipo di grafi ci\`o che interessa per l'analisi delle coppie di lettere scambiate sono i \emph{loop}, ossia le sequenze di lettere che si avvolgono su se stesse, come ad esempio \{\cypher{E}, \cypher{W}, \cypher{R}, \cypher{S}, \cypher{O}\} o \{\cypher{E}, \cypher{A}, \cypher{I}, \cypher{T}\}. Infatti, dall'opportuna analisi di ciascun \emph{loop} \`e possibile verificare se lo scambio di alcune coppie di lettere del \emph{Steckerverbindungen} sia compatibile con il \emph{crib} considerato ed escludere gli scambi di coppie che condurrebbero ad inconsistenze con le propriet\`a di \emph{non-identit\`a} e \emph{simmetria} della cifratura (cfr. paragrafo~\ref{sec:simply} per tali propriet\`a e~\cite{EnigmaSpec2,EnigmaSpec3} per il modo in cui correlare i \emph{loop} con tali propriet\`a). 
Alan Turing si rese conto per primo della possibilit\`a di rappresentare i \emph{loop} con dei circuiti elettrici, automatizzando l'analisi esaustiva delle coppie di lettere del \emph{Steckerverbindungen} per trovare quelle consistenti con il \emph{crib}. In seguito, il matematico Gordon Welchman trov\`o il modo di generalizzare l'approccio anche in assenza di \emph{loop}, permettendo di sfruttare ancor pi\`u efficientemente i \emph{crib}, sia per identificare le coppie di lettere scambiate dalla \emph{plug board} che per testare le possibili scelte dei rotori.

A met\`a del 1941, grazie alla combinazione delle tecniche e delle tecnologie sviluppate a Bletchley Park, le macchine \emph{bomba} erano finalmente in grado di trovare i settaggi giornalieri utilizzati dai tedeschi fornendo informazioni preziosissime ai servizi segreti Alleati.

\subsection{Un accenno al caso generale}\label{sec:generale}
Come detto la cifratura dei messaggi della Marina tedesca tramite Enigma era pi\`u complessa della versione usata da Esercito e Aviazione:
\begin{itemize}
\item I tre rotori venivano scelti tra 8 possibili rotori e non solo 5
\item I rotori aggiuntivi scattavano tutti e tre nella medesima posizione (la stessa del rotore \texttt{5})
\item I rotori aggiuntivi scattavano ogni 13 caratteri e non ogni 26
\item Le chiavi del messaggio non venivano inserite all'inizio del messaggio, ma generate a partire dai \emph{Kenngruppen} giornalieri con una complessa procedura (da fine 1941)
\end{itemize}

Non ci soffermeremo qui sugli adattamenti delle tecniche descritte alla decifrazione dei messaggi della Marina, perch\'e tali adattamenti sono non banali e vanno oltre lo scopo di questo studio non facendo direttamente uso del teorema di Bayes (il lettore interessato pu\`o trovare spiegazioni e dettagli in~\cite{Hosgood}). Menzioniamo solo il fatto che anche il calcolo del peso delle evidenze (che era stato introdotto nel paragrafo~\ref{sec:banburismus} e che verr\`a descritto in maggior dettaglio nella sezione~\ref{sec:FocusBayes}) risultava pi\`u complesso nel caso dei messaggi della Marina: siccome si potevano avere sia rotori che scattavano ogni $13$ lettere che rotori che scattavano ogni $26$ lettere, i messaggi cifrati che pure fossero stati \emph{in depth} si sarebbero disallineati allo scatto ``asincrono'' del rotore II e l'analisi dei match tra lettere uguali in posizioni uguali sarebbero risultati falsati da l\`\i{} in poi. Di conseguenza i coefficienti utilizzati nel calcolo del fattore di Bayes-Turing venivano opportunamente modificati per l'analisi di messaggi trasmessi dalla \emph{Kriegsmarine}.

\section{Focus sull'applicazione e il ruolo del teorema di Bayes}\label{sec:FocusBayes}

\subsection{Il teorema di Bayes}\label{sec:Bayes}
Il teorema di Bayes fu descritto per primo dall'inglese Richard Price nel 1763 sulla base degli scritti del revendo Thomas Bayes, suo amico morto nel 1761. Esso fornisce una formula di importanza cruciale per compiere correttamente il processo di inferenza, ossia per passare dalla probabilit\`a degli effetti rilevati in un fenomeno alla probabilit\`a delle cause che li hanno prodotti, e per aggiornare le valutazioni di probabilit\`a su un evento di interesse ogni qualvolta che nuove informazioni diventano disponibili. Pi\`u precisamente dati due eventi $A$ e $B$, il teorema di Bayes non \`e altro che la seguente formula
$$
P(A|B)\,=\,{ P(B|A) \,P(A) \over P(B)}
$$
che lega la probabilit\`a condizionata $A|B$, ossia la probabilit\`a che accada $A$ sapendo che si \`e osservato l'evento $B$, con la probabilit\`a condizionata $B|A$, ossia la probabilit\`a che accada $B$ sapendo che si \`e osservato l'evento $A$. Questa ``inversione'' di cause ed effetti osservati \`e quello che si chiama processo di inferenza perch\'e permette di valutare le probabilit\`a dell'evento $A$, che non conosciamo se sia accaduto o meno, sulla base del solo evento $B$ che si \`e osservato. Si pensi ad esempio al caso $B\ =$ ``sintomi osservati in un paziente'' e $A\ =$ ``malattia che causa i sintomi B''. Il teorema di Bayes rappresenta allora uno strumento ideale per fornire una diagnosi medica, tra tutte le malattie $A$ che possano potenzialmente produrre i sintomi $B$: conoscendo la probabilit\`a $P(A)$ di diffusione della malattia $A$ (nel territorio in cui opera il medico), la probabilit\`a $P(B)$ dei sintomi osservati tra tutte le possibili malattie che li possono procurare, e la \emph{verosimiglianza} o \emph{likelihood} $P(B|A)$, ossia quanto siano probabili i sintomi $B$ sapendo che effettivamente la malattia $A$ \`e presente, diventa immediato valutare la probabilit\`a $P(A|B)$ che sia in effetti la malattia $A$ a causare i sintomi osservati $B$.

Come accennato, inoltre, la formula di Bayes gioca anche un ruolo chiave nell'aggiornamento delle nostre credenze quando nuove informazioni/osservazioni diventano disponibili. Quando si esce dal dominio della probabilit\`a che viene insegnata alle scuole secondarie e si ha a che fare con processi pi\`u complessi del lancio di una moneta o di un dado ideali (ossia perfettamente equi), infatti, risulta alquanto difficile se non impossibile assegnare una valutazione di probabilit\`a del tipo ``rapporto tra numero dei casi favorevoli e numero dei casi possibili''. Diventa invece molto efficace affidarsi all'approccio \emph{soggettivista} alla probabilit\`a, sulle orme di Bruno de Finetti~\cite{DeFinetti} e, nel mondo anglosassone, di Leonard J. Savage~\cite{Savage}, secondo il quale la valutazione delle probabilit\`a di un evento $A$ dipender\`a dalle informazioni a disposizione di colui che la deve assegnare e sar\`a pari alla quota che saremmo disposti a pagare per partecipare ad una lotteria in cui si vinca se l'evento $A$ si realizza. Si noti che in questo caso, ``soggettivo'' non significa ``arbitrario'' ma semplicemente ``influenzato dalle informazioni a disposizione del soggetto valutatore'': la probabilit\`a assegnata ad un qualunque evento $A$ non pu\`o che dipendere dalle informazioni a nostra disposizione, in quanto diverse informazioni produrranno diverse valutazioni, ma al contempo individui diversi aventi le medesime informazioni giungeranno alla medesima valutazione. Si pensi ad esempio al seguente esperimento: 
\begin{itemize}
\item si sceglie una moneta da un sacchetto che contiene un egual numero di monete eque e di monete false dotate di due teste; 
\item senza guardare se la moneta abbia due teste o solo una, la si lanci e se ne osservi il risultato; 
\item se il risultato \`e testa, quale sar\`a la probabilit\`a che si tratti di una moneta equa?
\item se si lancia nuovamente la moneta ed il risultato \`e ancora testa, quale sar\`a la probabilit\`a che si tratti di una moneta equa?
\item e si lancia la stessa moneta dieci volte e tutte e dieci le volte esce testa, quale sar\`a la probabilit\`a che si tratti di una moneta equa?
\end{itemize}

\`E abbastanza intuitivo il fatto che ad ogni lancio successivo in cui osserviamo l'esito \emph{testa} (\coin{T}) si tenda a propendere maggiormente verso l'opzione ``moneta falsa'' piuttosto che verso l'opzione ``moneta vera'', visto che ci si aspetta che se la moneta fosse vera qualche esito \emph{croce} (\coin{C}) dovrebbe essere osservato durante i lanci successivi (ovviamente un singolo esito \coin{C} da un lancio ci dar\`a la certezza che la moneta sia vera). Tuttavia il modo corretto per valutare questa probabilit\`a in termini \emph{quantitativi} \`e tutt'altro che banale.

\n Infatti \`e fondamentale osservare che non viene chiesto quale sia la probabilit\`a di ottenere \coin{T} dopo un lancio o dieci \coin{T} in dieci lanci, sapendo che la moneta \`e equa (o falsa). Se queste fossero le domande poste, le corrette risposte sarebbero immediate: per una moneta falsa, la probabilit\`a di osservare \coin{T} \`e $1$ sia dopo un lancio che dopo dieci lanci; per una moneta equa, la probabilit\`a di dare \coin{T} al primo lancio \`e naturalmente $1/2$, e quella di dare $n$ volte \coin{T} consecutivamente in $n$ lanci \`e $(1/2)^n$. Qui per\`o siamo di fronte ad un quesito differente: noi non sappiamo se la moneta sia equa o falsa e quindi dobbiamo ricorrere al teorema di Bayes per invertire correttamente le probabilit\`a di interesse!

Indicando con $A$ l'evento ``la moneta \`e equa'' e con $B$ l'evento ``osservare \coin{T} al primo lancio'', avremo che la verosimiglianza $P(B|A)$ \`e pari a $0.5$ e che la probabilit\`a $P(B)$ pu\`o essere calcolata applicando il cosiddetto \emph{teorema della probabilit\`a assoluta} (o \emph{delle probabilit\`a totali}) che discende direttamente dagli assiomi di Kolmogorov per la probabilit\`a:
$$
P(B) = P(B|A)~\cdot~ P(A) + P(B| \mbox{non--}A)~\cdot~ P(\mbox{non--}A) = 0.5~\cdot~ 0.5 + 1~\cdot~ 0.5 = 3/4
$$
dove $P(A)= P(\mbox{non--}A)=0.5$ perch\'e abbiamo detto che la moneta era estratta da un sacchetto in cui monete eque e monete false erano in egual numero e $P(B|\mbox{non--}A)=1$ perch\'e se la moneta \`e falsa non si pu\`o che osservare l'esito \coin{T}. Qui $P(A)$ \`e la cosiddetta valutazione \emph{a priori} dell'evento $A$, in quanto \`e la probabilit\`a che gli assegnamo prima di osservare alcun lancio della moneta estratta.

\n Applicando il teorema di Bayes, la probabilit\`a aggiornata, o probabilit\`a \emph{a posteriori}, $P(A|B)$ sar\`a quindi data da $(1/2~\cdot~ 1/2) / (3/4) = 1/3$. Ossia, l'aver osservato \coin{T} dopo il primo lancio ha modificato la nostra valutazione che la moneta fosse equa facendola calare da $1/2$ a $1/3$. 
Dopo il secondo lancio possiamo applicare nuovamente il teorema di Bayes, utilizzando stavolta come probabilit\`a a priori quelle ottenute dopo il primo lancio, ossia $P(A) = 1/3$ e $P(\mbox{non--}A)=2/3$. Avremo quindi
$$
P(B) = P(B|A)~\cdot~ P(A) + P(B| \mbox{non--}A)~\cdot~ P(\mbox{non--}A) = 0.5~\cdot~ 0.\overline{3} + 1~\cdot~ 0.\overline{6} = 5/6
$$
e $P(A|B) = (1/2~\cdot~ 1/3) / (5/6) = 1/5$. E cos\`\i{} via, ad ogni ulteriore lancio della moneta il cui esito \`e \coin{T}, la nostra valutazione dell'equit\`a della moneta si ridurr\`a fino ad essere inferiore all'$1/1000$ dopo i dieci esiti \coin{T} di fila.

Le valutazioni della probabilit\`a ottenute applicando il teorema di Bayes sono quelle matematicamente pi\`u corrette che si possono offrire sulla base delle informazioni a disposizione del valutatore. \`E tuttavia ovvio, come detto, che un altro valutatore, avente diverse informazioni a disposizione, fornirebbe una diversa stima: ad esempio se un altro partecipante all'esperimento guardasse la moneta estratta da entrambi i lati, potrebbe immediatamente valutare come $0$ o $1$ la probabilit\`a che la moneta sia equa! Entrambe le valutazioni della probabilit\`a, quella di colui che ha solo a disposizione gli esiti dei lanci ripetuti e quella di colui che ha guardato entrambe le facce della moneta, sono ``corrette'', ovvero razionali, in quanto sono consistenti con le informazioni a disposizione. \`E interessante osservare che, anche quando si parta da diverse valutazioni a priori della probabilit\`a di un evento, gli aggiornamenti delle valutazioni a seguito di nuove osservazioni $B$ avvengono utilizzando un medesimo fattore (il quoziente della verosimiglianza $P(B|A)$ e della probabilit\`a $P(B)$), e quindi variano in maniera assolutamente coerente, crescendo o decrescendo entrambe alla luce di osservazioni $B$ a favore o contrarie alla valutazione dell'evento $A$. L'algoritmo di aggiornamento, in altre parole, \`e completamente \emph{oggettivo}.

Concludiamo questa breve panoramica sul teorema di Bayes menzionandone una formulazione alternativa che verr\`a utilizzata nel seguito. Invece di valutare la probabilit\`a di un evento $A$, si \`e spesso interessati a valutare le \emph{odd} relative ad $A$, ossia il valore del rapporto  tra la probabilit\`a che un evento $A$ accada e la probabilit\`a che tale evento non accada (in formula si ha ${\cal O}(A)=P(A) / P(\mbox{non--$A$})$): riformulando l'espressione data sopra in termini di \emph{odd} avremo
\begin{align*}
{\cal O}(A|B)\,&=\,{ P(A|B) \over P(\mbox{non--$A$}|B)}\,=\,{ P(B|A) \,P(A) \over P(B|\mbox{non--$A$}) \,P(\mbox{non--$A$})}\\
&~~~~~~~~~\,=\,{ P(B|A) \over P(B|\mbox{non--$A$}) }\, {\cal O}(A)
\end{align*}
in cui si pu\`o vedere come l'aggiornamento delle \emph{odd} passi attraverso la moltiplicazione per il cosiddetto \emph{fattore di Bayes-Turing} che altro non \`e se non il rapporto tra le verosimiglianze di $B$ dati rispettivamente gli eventi $A$ e $\mbox{non--}A$.

\subsection{Il teorema di Bayes nel \emph{Banburismus}}\label{sec:BayesBanburismus}
Andiamo ora a focalizzare la nostra attenzione sull'utilizzo del teorema di Bayes nel processo denominato \emph{Banburismus}. Come accennato nella sezione precedente, il contributo principale fornito dal teorema consiste nel calcolo del fattore di Bayes-Turing, che permette di classificare le diverse possibili posizioni delle coppie di messaggi cos\`\i{} da identificare quelle che pi\`u verosimilmente rappresentavano due messaggi \emph{in depth}. Ricordiamo che per coppie di messaggi \emph{in depth} si intendono coppie in cui la cifratura delle lettere in entrambi i messaggi procede in parallelo perch\'e \`e come se le macchine Enigma fossero partite con i rotori nelle stesse posizioni. 

La prima osservazione necessaria per comprendere come i crittoanalisti di Bletchley Park giunsero all'idea di utilizzare il fattore di Bayes-Turing per attaccare la cifratura dei messaggi tedeschi \`e la seguente: nei messaggi di qualunque lingua le lettere non appaiono uniformemente distribuite, come invece accadrebbe se si trattasse di sequenze casuali estratte da un alfabeto con un campionamento uniforme. Per fissare le idee riportiamo nelle seguenti tabelle la frequenza di occorrenza delle lettere principali in Italiano e in Inglese.
\begin{center}
\begin{tabular}{|c|l|}
\hline
\multicolumn{2}{|c|}{Italiano}\\
\hline
\hline
\cypher{E} & $11,79 \%$\\
\cypher{A} & $11,74 \%$\\
\cypher{I} & $11,28 \%$\\
\cypher{O} & $9,83 \%$\\
\cypher{N} & $6,88 \%$\\
\hline
\end{tabular}
~~~~~~~~~~~~
\begin{tabular}{|c|l|}
\hline
\multicolumn{2}{|c|}{Inglese} \\
\hline
\hline
\cypher{E} & $12,31 \%$\\
\cypher{T} & $9,59 \%$\\
\cypher{A} & $8,05 \%$\\
\cypher{O} & $7,94 \%$\\
\cypher{N} & $7,19 \%$\\
\hline
\end{tabular}
\end{center}
Una conseguenza immediate di questo fatto \`e che, se noi sovrapponiamo due messaggi di senso compiuto, la probabilit\`a di trovare una stessa lettera nella medesima posizione nei due messaggi \`e maggiore di quella che si avrebbe sovrapponendo due sequenze di lettere casuali.
Infatti, se noi sovrapponiamo due sequenze assolutamente casuali di lettere ci aspettiamo di trovare lettere uguali in posti uguali con una probabilit\`a di $1/26$. Se invece andiamo a sovrapporre due frasi di senso compiuto possiamo trovare un numero maggiore di lettere uguali nella stessa posizione!

Andiamo ad analizzare un esempio, per semplicit\`a: confrontando le seguenti frasi lunghe 33 lettere otteniamo che
$$
\begin{array}{c}
\mbox{\cypher{R I \textred{P} \textred{R} O V A T E Q U \textred{E} S T O E S P E R I M E N T O A L T R O V E}}\\
\mbox{\cypher{H O \textred{P} \textred{R} E S O U N A S \textred{E} Q U E N Z A Q U A S I C A S U A L E Q U I}}
\end{array}
$$
si hanno  $3$ ``match'' su una stringa di $33$ lettere, quindi un rapporto di $1/11$ che \`e maggiore di $1/26$. E cos\`\i{} accade in maniera ``generica'', ossia questa \`e la norma e non una eccezione dovuta alle particolari stringhe scelte.

Come ci aiuta questa conoscenza con il problema dell'identificazione dei rotori? Il punto chiave \`e che la frequenza dei match resta la stessa anche se i messaggi sono cifrati (purch\'e lo siano con una medesima procedura): ovviamente le varie lettere saranno cambiate in altre, ma le lettere cifrate saranno le medesime in posizioni coincidenti con una frequenza maggiore se siamo partiti da frasi di senso compiuto piuttosto che se siamo partiti da sequenze alfabetiche casuali. Questo significa che se fossimo in grado di ``riallineare'' il terzo rotore, in modo da avere messaggi \emph{in depth}, da quel punto in poi la frequenza con cui incontriamo una medesima lettera nella medesima posizione in entrambi i messaggi dovrebbe distribuirsi come quella riscontrabile in messaggi in chiaro.

L'idea di Turing e del suo gruppo fu quindi quella di confrontare tutte le possibili 51 posizioni relative dei due messaggi e classificarne la verosimiglianza di essere una posizione \emph{in depth}, sulla base di uno score definito a partire dal numero di lettere uguali (o blocchi di lettere) nella medesima posizione. Si noti che le posizioni vanno da $-26$ (ossia $26$ posizioni a sinistra) fino a $+26$ (ossia $26$ posizioni a destra), ma che la posizione $0$ sappiamo gi\`a non poter essere corretta: se i due messaggi fossero \emph{in depth} dopo uno spostamento di $0$ posizioni, vorrebbe dire che sono \emph{in depth} gi\`a originalmente, ma questo non \`e possibile perch\'e la terza lettera della chiave \`e diversa, e quindi i due messaggi posseggono uno \emph{shift} non nullo.
Presa quindi una coppia di messaggi in cui solo la terza lettera della chiave specifica del messaggio \`e differente, ossia una coppia di messaggi come la seguente:
$$
\begin{array}{l}
\mbox{\cypher{"VFG"  :  GXCYBGDSLVWBDJLKWIPEHVYGQZWDTHRQXIKEESQSSPZXARIXEABQIRUCKHGWUEBPF}}\\
\mbox{\cypher{"VFX"  :  YNSCFCCPVIPEMSGIZWFLHESCIYSPVRXMCFQAXVXDVUQILBJUABNLKMKDJMENUNQ}}
\end{array}
$$
procediamo tendendo fisso uno dei due messaggi e facendo scorrere l'altro a destra e sinistra:
$$
\begin{array}{l}
\mbox{\cypher{\fbox{GXCYBGDSLVWBDJLKWIPEHVYGQZWDTHRQXIKEESQSSPZXARIXEABQIRUCKHGWUEBPF}}}\\
\longrightarrow\mbox{\cypher{\phantom{YBGDSLVWB}\,YNSCFCCPVIPEMSGIZWFLHESCIYSPVRXMCFQAXVXDVUQILBJUABNLKMKDJMENUNQ}}
\end{array}
$$
$$
\begin{array}{l}
\phantom{\longleftarrow}\mbox{\cypher{\phantom{YNSCFC}\fbox{GXCYBGDSLVWBDJLKWIPEHVYGQZWDTHRQXIKEESQSSPZXARIXEABQIRUCKHGWUEBPF}}}\\
\mbox{\cypher{YNSCFCCPVIPEMSGIZWFLHESCIYSPVRXMCFQAXVXDVUQILBJUABNLKMKDJMENUNQ\phantom{IRUCKHGWU\!}}}\longleftarrow
\end{array}
$$

\n Per ciascuna posizione vediamo se le ripetizioni di lettere uguali assomiglia a quella di una stringa di caratteri casuali, o a quella di un messaggio!

Ad esempio, spostando il secondo messaggio di 8 posizioni a destra troviamo 1 gruppo di tre lettere sovrapposto:
$$
\begin{array}{l}
\mbox{\cypher{GXCYBGDSLVWBDJLKWIPEHVYGQZWDTHRQXIKEESQSSPZXARIXEABQIRUCKHGWUEBPF}}\\
\mbox{\cypher{\phantom{GXCYBGDS}YNSCFCCPVIPEMSGIZWFLHESCIYSPVRXMCFQAXVXDVUQILBJUABNLKMKDJMENUNQ}}\\
\mbox{\cypher{\phantom{GXCYBGDSYNSCFCCPV}---}}
\end{array}
$$
Spostando invece il secondo messaggio di 9 posizioni a destra troviamo 5 lettere singole sovrapposte e 2 gruppi da due lettere sovrapposte:
$$
\begin{array}{l}
\mbox{\cypher{GXCYBGDSLVWBDJLKWIPEHVYGQZWDTHRQXIKEESQSSPZXARIXEABQIRUCKHGWUEBPF}}\\
\mbox{\cypher{\phantom{GXCYBGDSL}YNSCFCCPVIPEMSGIZWFLHESCIYSPVRXMCFQAXVXDVUQILBJUABNLKMKDJMENUNQ}}\\
\mbox{\cypher{\phantom{GXCYBGDSLVWBDJLKWIPEHVY}-\phantom{Q}--\phantom{DT}-\phantom{RQX}-\phantom{KEESQSSPZX}-\phantom{RI}-\phantom{EAB}--}}
\end{array}
$$
 
\n Come possiamo ora confrontare i due risultati (e gli altri 49) per ottenere un ranking che ci permetta di classificare ciascuna posizione come potenzialmente \emph{in depth} oppure no? Con il teorema di Bayes!

Nel nostro caso, infatti, possiamo pensare che l'evento a cui siamo interessati sia $A\ =$ ``i messaggi sono \emph{in depth} con quello \emph{shift}'' e che le evidenze che ci permettono di aggiornare la nostra valutazione siano $B\ =$ ``abbiamo osservato un certo numero di lettere uguali tra i due messaggi''. A questo punto vogliamo valutare le \emph{odd} che in effetti valga $(A|B)$, ossia che in effetti i messaggi siano \emph{in depth}. Ricordando la formula data nel paragrafo~\ref{sec:Bayes} per il fattore, e indicando con $N$ il numero di lettere effettivamente sovrapposte tra i due messaggi dopo lo \emph{shift} e con $M$ il numero di lettere effettivamente coincidenti, otterremo:
\begin{align*}
{\cal O}(\mbox{Posiz. corretta}\,&|\,\mbox{$M$ match su $N$ overlap})\\ 
&\,=\,
{P(\mbox{Posiz. corretta}\,|\,\mbox{$M$ match su $N$ overlap})
\over
P(\mbox{Posiz. sbagliata}\,|\,\mbox{$M$ match su $N$ overlap})
} \\
&\,=\,
{P(\mbox{$M$ match su $N$ overlap}\,|\,\mbox{Posiz. corretta})\cdot P(\mbox{Posiz. corretta})
\over
P(\mbox{$M$ match su $N$ overlap}\,|\,\mbox{Posiz. sbagliata})
\cdot P(\mbox{Posiz. sbagliata})
} \\
&\,=\,
{P(\mbox{$M$ match su $N$ overlap}\,|\,\mbox{Posiz. corretta})\over
P(\mbox{$M$ match su $N$ overlap}\,|\,\mbox{Posiz. sbagliata})
}\,\cdot\,{\cal O}(\mbox{Posiz. corretta})
\end{align*}

La nostra attenzione si focalizza qui sul fattore di Bayes-Turing, ossia al fattore moltiplicativo che permette di passare dalle \emph{odd a priori} a quelle aggiornate: se il fattore \`e maggiore di $1$, l'evidenza ``$M$ match su $N$ overlap'' fa crescere le \emph{odd} e quindi si tratta di un'evidenza a favore; se il fattore \`e minore di $1$, l'evidenza fa decrescere le \emph{odd} e quindi si tratta di un'evidenza contraria.

\n Possiamo quindi classificare le varie situazioni di \emph{overlap} in base al fattore ottenuto, stilando un \emph{ranking} di quali siano maggiormente credibili come corrispondenti a due messaggi \emph{in depth}. Si noti anche che nella composizione di tale \emph{ranking} non \`e necessario assegnare le \emph{odd} a priori, perch\'e non si \`e interessati direttamente al valore numerico della quota aggiornata ma solo al fattore moltiplicativo.

\n Il calcolo del fattore pu\`o essere completato esplicitamente: 
\begin{align*}
{P(\mbox{$M$ match su $N$ overlap}\,|\,\mbox{Posiz. corretta})
\over
P(\mbox{$M$ match su $N$ overlap}\,|\,\mbox{Posiz. sbagliata})
} &\,=\, 
{\left({1\over 17}\right)^M\cdot \left({16\over 17}\right)^{N-M}
\over
\left({1\over 26}\right)^M\cdot \left({25\over 26}\right)^{N-M}
}
\end{align*}
dove $1/17$ \`e la probabilit\`a di avere match di lettere in messaggi di tipo militare in Tedesco, frutto di una serie di analisi statistiche effettuate a cavallo delle due guerre mondiali. Nel seguito chiameremo \emph{peso delle evidenze} $W$($M$ match su $N$ overlap) il logaritmo del fattore di Bayes-Turing, che risulta pi\`u facilmente calcolabile rispetto al numero originale del fattore.

\n Al numero $W$($M$ match su $N$ overlap) appena trovato, vanno in realt\`a fatte delle correzioni (additive): 
\begin{itemize}
\item Un \emph{Bonus} se si trovano sequenze di match consecutivi, perch\'e rappresentano il match di una sillaba o di parte di essa sui due messaggi.
\item Un \emph{Malus} che cresce al crescere del numero di posizioni per cui si \`e shiftata la sequenza, perch\'e diventa pi\`u probabile l'avvenire di uno scatto del rotore intermedio e che i match successivi siano frutto di coincidenza.
\end{itemize}
La formula finale per il \emph{peso delle evidenze} utilizzato a Bletchley Park era dunque il seguente
$$
{\cal W}(\mbox{$M$ match su $N$ overlap})\,=\,
\log {\left({1\over 17}\right)^M\cdot \left({16\over 17}\right)^{N-M}
\over
\left({1\over 26}\right)^M\cdot \left({25\over 26}\right)^{N-M}}
\,+\, \mbox{Bonus} \,-\,  \mbox{Malus}
$$
dove i valori numerici utilizzati per $\mbox{Bonus}$ e $\mbox{Malus}$ venivano presi da opportune tavole numeriche (alcuni esempi di tali tavole possono essere trovare in~\cite{Hosgood}, mentre la loro descrizione completa \`e presente in~\cite{EnigmaSpec1}).

Tramite questa formula possiamo quindi classificare tutte le posizioni in base al loro \emph{peso} e scegliere quella (o quelle) con peso maggiore come pi\`u probabilmente corretta.

\n Nell'esempio precedente dei messaggi cifrati con \cypher{VFG} e \cypher{VFX}, la posizione $+9$ era proprio quella con un peso maggiore:
$$
\begin{array}{l}
\mbox{\cypher{GXCYBGDSLVWBDJLKWIPEHVYGQZWDTHRQXIKEESQSSPZXARIXEABQIRUCKHGWUEBPF}}\\
\mbox{\cypher{\phantom{GXCYBGDSL}YNSCFCCPVIPEMSGIZWFLHESCIYSPVRXMCFQAXVXDVUQILBJUABNLKMKDJMENUNQ}}\\
\mbox{\cypher{\phantom{GXCYBGDSLVWBDJLKWIPEHVY}-\phantom{Q}--\phantom{DT}-\phantom{RQX}-\phantom{KEESQSSPZX}-\phantom{RI}-\phantom{EAB}--}}
\end{array}
$$

\n Dalla sovrapposizione avevamo che le lettere coincidenti erano una sequenza di 1 lettera, poi 2 lettere, poi 1 lettera per quattro volte e ancora 2 lettere, per un totale di $M=9$ match su $N=56$ lettere di overlap. Calcolando il logaritmo del fattore di Bayes-Turing ed aggiungendo gli opportuni $\mbox{Bonus}$ \& $\mbox{Malus}$, concluderemo allora che $X = G + 9$ con peso \odds{5:1}. Nella tabella successiva abbiamo elencato per ogni \emph{shift} possibile la sequenza di effettive lettere di match $M$ rispetto al numero di lettere di overlap $N$.

\begin{center}
\begin{tabular}{|l|l|l||l|l|l|}
\hline
SHIFT & SEQUENZA MATCH & OVERLAP & SHIFT & SEQUENZA MATCH & OVERLAP\\
\hline
\hline
-25 & - & 38 & 1 & 1, 1, 2 & 63\\
-24 & 1 & 39 & 2 & 1, 1, 1 & 63\\
-23 & 1 & 40 & 3 & 1, 1, 1 & 62\\
-22 & 1, 1 & 41 & 4 & 1, 1 & 61\\
-21 & 1 & 42 & 5 & 1, 1, 1 & 60\\
-20 & 1 & 43 & 6 & - & 59\\
-19 & 1, 1, 1 & 44 & 7 & 1, 1 & 58\\
-18 & 1, 1 & 45 & 8 & 3 & 57\\
-17 & - & 46 & 9 & 1, 2, 1, 1, 1, 1, 2 & 56\\
-16 & 1 & 47 & 10 & - & 55\\
-15 & 1 & 48 & 11 & 1, 1, 1 & 54\\
-14 & 1 & 49 & 12 & - & 53\\
-13 & - & 50 & 13 & 1, 1, 1, 1, 1, 1 & 52\\
-12 & 1 & 51 & 14 & 1, 2, 1 & 51\\
-11 & 1, 1, 1 & 52 & 15 & 2 & 50\\
-10 & 1, 1, 1 & 53 & 16 & 1 & 49\\
-9 & 1, 1 & 54 & 17 & 1, 1, 1, 1 & 48\\
-8 & - & 55 & 18 & 1, 1 & 47\\
-7 & 1, 1, 1 & 56 & 19 & 1 & 46\\
-6 & 1, 1, 1 & 57 & 20 & - & 45\\
-5 & 1 & 58 & 21 & - & 44\\
-4 & 1, 1, 1, 1 & 59 & 22 & 1, 1 & 43\\
-3 & 1, 1, 1 & 60 & 23 & 1 & 42\\
-2 & 1 & 61 & 24 & 1, 1, 1, 1 & 41\\
-1 & 1, 1 & 62 & 25 & 1 & 40\\
\hline
\end{tabular}
\end{center}

\n Omettiamo il calcolo dei pesi delle evidenze per ciascuna dei possibili \emph{shift}: essi possono essere trovati applicando le formule in~\cite{Hosgood}.

Menzioniamo invece il fatto che Turing svilupp\`o l'approccio basato sull'uso del \emph{peso delle evidenze} in varie direzioni, riuscendo ad applicarlo allo studio di \emph{toy model} crittografici~\cite{Turing,TuringComm}, all'attacco verso altre macchine cifranti~\cite{SZ42,Tunny} e anche alla valutazione dell'attendibilit\`a dei \emph{crib} (di cui abbiamo parlato brevemente nel paragrafo~\ref{sec:crib}).

\section{Conclusioni}\label{sec:conclusioni}

In questo scritto abbiamo cercato di fornire uno sguardo particolare sul problema della decifrazione dei messaggi tedeschi durante la Seconda Guerra Mondiale, focalizzando la nostra attenzione sull'utilizzo da parte di  Turing del teorema di Bayes (per ridurre il carico di lavoro sulle macchine \emph{bomba} che avevano il compito di individuare la chiave di cifratura utilizzata giornalmente per le macchine Enigma). 

Da un punto di vista tecnico, il contributo dato da tale teorema non risulta forse dirompente quanto la costruzione stessa delle macchine \emph{bomba}, che costituirono un enorme balzo tecnico e tecnologico rispetto all'idea stessa di calcolatore che si aveva al tempo. \`E tuttavia innegabile che senza il contributo del fattore di Bayes-Turing il lavoro delle macchine sarebbe risultato proibitivo, essendo troppo elevato il numero di configurazioni da analizzare e testare, e forse molte delle informazioni di intelligence che si rivelarono fondamentali durante il conflitto non sarebbero mai state decifrate in tempo utile.

Questo aspetto \`e stato tipicamente poco esplorato ed \`e per questa ragione che abbiamo voluto cominciare a raccontarlo, pur trattandosi di un aspetto alquanto tecnico che quindi richiede un notevole sforzo di approfondimento al lettore. Purtoppo la comprensione dell'uso del teorema di Bayes richiedeva la comprensione delle regole ``base'' della cifratura tramite Enigma, e queste a loro volta richiedevano la descrizione del funzionamento meccanico delle macchine cifranti. Ci auguriamo comunque che almeno i punti fondamentali della trattazione siano risultati comprensibili, anche per un lettore non precedentemente esperto sul tema della crittografia e delle macchine cifranti del XX secolo.

\subsection*{Ringraziamenti}

L'idea di questo articolo \`e stata concepita e sviluppata durante i seminari della \emph{Decision Making School} di Roma, finanziata da MBDA. Gli autori vogliono ringraziare i professori Giulio D'Agostini dell'Universit\`a di Roma ``Sapienza'' e Julia Mortera dell'Universit\`a di Roma Tre per le fruttuose discussioni e per i consigli sulla presentazione del tema.


\begin{thebibliography}{99}
\bibitem{LIGO1} Benjamin P. Abbott et al. (LIGO Scientific Collaboration and Virgo Collaboration), \emph{Observation of Gravitational Waves from a Binary Black Hole Merger}, Phys. Rev. Lett. {\bf 116} (2016), 061102.
\bibitem{LIGO2} Benjamin P. Abbott et al. (LIGO Scientific Collaboration and Virgo Collaboration), \emph{Properties of the binary black hole merger GW150914}, Phys. Rev. Lett. {\bf 116} (2016), 241102.
\bibitem{EnigmaSpec1} C. Hugh O'D. Alexander, Cryptographic history of work on German Naval ENIGMA, Public Record Office, Kew, Surrey, UK, 1946.
\bibitem{Bellaso} Giovan Battista Bellaso, La Cifra del Sig. Giovan Battista Bellaso. Venezia, 1553.
\bibitem{Churchill} Winston Churchill, The World Crisis 1911-1918, Barnes and Noble, 1993, ISBN 9781566191883.
\bibitem{NavyWWI} Julian S. Corbett, History Of The Great War - Naval Operations, Volume 3, Spring 1915 to June 1916. Longmans Green And Co., London, 1923.
\bibitem{SZ42} Donald W. Davies, \emph{The Lorenz Cipher Machine SZ42}, Cryptologia {\bf 19} (1995), no. 1, 39--61.
\bibitem{DeFinetti} Bruno DeFinetti, \emph{Sul significato soggettivo della probabilit\`a}, in Fundamenta Mathematicae, {\bf XVII} (1931), 298--329.
\bibitem{UltraRef1} Harold C. Deutsch, \emph{The Historical Impact of Revealing The Ultra Secret}, NSA Cryptologic Spectrum Articles {\bf 8} (1978), no. 1.
\bibitem{NASA1} Homayoon Dezfuli, Dana Kelly, Curtis Smith, Kurt Vedros \& William Galyean, \emph{Bayesian Inference for NASA Probabilistic Risk and Reliability Analysis}, NASA Risk Docs {\bf NASA/SP-2009-569}. [available online: \url{http://www.hq.nasa.gov/office/codeq/doctree/SP2009569.pdf}]
\bibitem{EnigmaSpec2} Graham Ellsbury, The Turing Bombe, 1998. [available online: \url{http://www.ellsbury.com/bombe1.htm}]
\bibitem{Erodoto} Erodoto, Le Storie. UTET, Torino, 2014, ISBN 9788851122294.
\bibitem{Garbolino} Paolo Garbolino, Probabilit\`a e logica della prova. Giuffr\`e Editore, 2014, ISBN 9788814189630.
\bibitem{Gellio} Aulo Gellio, Notti Attiche. UTET, Torino, 2013, ISBN 9788841893050
\bibitem{Hosgood} Steven Hosgood, \emph{All You Ever Wanted to Know About Banburismus but were Afraid to Ask}, 2007. [available online: \url{http://stoneship.org.uk/~steve/banburismus.html}] 
\bibitem{Jefferson} Thomas Jefferson, The wheel cypher. Jefferson's Notes and Copy, Thomas Jefferson Papers, Library of Congress. [available online: \url{http://memory.loc.gov/cgi-bin/ampage?collId=mtj1\&fileName=mtj1page056.db\&recNum=47}]
\bibitem{Kasiski} Friedrich W. Kasiski, Die Geheimschriften und die Dechiffrir-Kunst. Mit besonderer Ber\"ucksichtigung der deutschen und der franz\"osischen Sprache. E.S. Mittler und Sohn, Berlino, 1863.
\bibitem{Plutarco} Plutarco, Vite parallele: Lisandro e Silla. BUR Rizzoli, Milano, 2001, ISBN 9788817128964.
\bibitem{Polibio} Polibio, Storie (Libri VII-XI - Volume quarto). A cura di Domenico Musti. BUR Rizzoli, Milano, 2002, ISBN 9788817100269.
\bibitem{EnigmaSpec3} Tony Sale, Virtual Wartime Bletchley Park, 2002. [available online: \url{https://www.codesandciphers.org.uk/virtualbp/tbombe/tbombe.htm}]
\bibitem{Savage} Leonard J. Savage, \emph{The Foundations of Statistics Reconsidered}, Proc. Fourth Berkeley Symp. on Math. Statist. and Prob. {\bf 1} (1961), 575--586.
\bibitem{Patent} Arthur Scherbius, US Patent for a Ciphering Machine, US1657411. [available online: \url{www.google.com/patents/US1657411}]
\bibitem{Shannon} Claude E. Shannon, \emph{Communication Theory of Secrecy Systems}, Bell System Technical Journal {\bf 28}, 656--715.
\bibitem{ClinicBayes} David J. Spiegelhalter, Keith R. Abrams \& Jonathan P. Myles, Bayesian Approaches to Clinical Trials and Health-Care Evaluation, Wiley, Hoboken NJ, 2004, ISBN 9780471499756.
\bibitem{NASA2} Michael Stamatelatos \& Homayoon Dezfuli, Probabilistic Risk Assessment Procedures Guide for NASA Managers and Practitioners, NASA Risk Docs NASA/SP-2011-3421. [available online: \url{http://www.hq.nasa.gov/office/codeq/doctree/SP20113421.pdf}]
\bibitem{Svetonio} Svetonio, Le vite dei dodici Cesari. UTET, Torino, 2009, ISBN 9788802080772.
\bibitem{Turing} Alan Turing, \emph{The Applications of Probability to Cryptography}, UK National Archives manuscript {\bf HW 25/37} [available online: \url{https://arxiv.org/abs/1505.04714}]
\bibitem{Vigenere} Blaise de Vigen\`ere, Traict\'e des chiffres ou secr\`etes mani\`eres d'escrire. Parigi, 1586.
\bibitem{Tunny} VV.AA., Breaking Teleprinter Ciphers at Bletchley Park: An edition of I.J. Good, D. Michie and G. Timms: General Report on Tunny with Emphasis on Statistical Methods (1945). Editors James A. Reeds, Whitfield Diffie, J. V. Field, 2012, Wiley, Hoboken NJ, ISBN 9780470465899.
\bibitem{UltraRef2} VV.AA., \emph{Ultra and the Battle of the Atlantic}, NSA Cryptologic Spectrum Articles {\bf 8} (1978), no. 1.
\bibitem{UltraRef3} Frederick Winterbotham, The Ultra Secret. Weidenfeld and Nicolson, London, 1974, ISBN 0297768328.
\bibitem{TuringComm} Sandy Zabell, \emph{Commentary on Alan M. Turing: The Applications of Probability to Cryptography}, Cryptologia,  {\bf  36} (2012), no. 3, 191--214.
\end{thebibliography}
\end{document}